\documentclass[12pt]{article}
 \usepackage{amsmath,amsthm,epsfig,amsfonts,amssymb}
 \usepackage[latin1]{inputenc}
 \usepackage{amstext}
 \usepackage{color}
 \usepackage{float}
 \usepackage{natbib}  
\newtheorem{proposition}{Proposition}
\newtheorem{theorem}{Theorem}
\newtheorem{remark}{Remark}
\newtheorem{Assumption}{Assumption}
\begin{document}
\begin{titlepage}
\begin{center}
{\Large Indifference Pricing of Defaultable Bond\\
with Stochastic Intensity Model  }\\
\vspace{2.5 cm}
R\'egis Houssou \protect\footnote{R\'egis Houssou, Institute of Mathematics, University of Neuch\^atel. E-mail: regis.houssou@unine.ch} and
Olivier Besson \protect\footnote{Olivier Besson, Institute of Mathematics, University of Neuch\^atel, Switzerland.}
\vspace{0.5cm}

{\today}
    
\begin{abstract}
The utility-based pricing of defaultable bonds in the case of stochastic intensity models of default risk is discussed. 
The Hamilton-Jacobi-Bellman (HJB) equations for the value functions is derived.
A finite difference method is used to solve this problem. 
The yield-spreads for both buyer and seller are extracted. The behaviour of the spread curve given the default intensity is analyzed. Finally the impacts of the risk aversion and the correlation coefficient are discussed.\\
\vspace{1cm}
\end{abstract}

\vspace{6cm}
\end{center}
\end{titlepage}

\section{Introduction}

Credit risk models can be partitioned into two groups known as structural models and reduced form models. \\
Structural models were pioneered by \cite{M}. The basic idea, common to all structural-type models, is that a company defaults on its debt if the value of the assets of the company falls below a certain default point. Such a default can be expected. In these models it has been demonstrated that default can be modelled as an option and, as a result, researchers were able to apply the same principles used for option pricing to the valuation of risky corporate securities. The application of option pricing avoids the use of risk premium and tries to use other marketable securities to price the option. The use of option pricing theory set forth by Merton offers much more accurate prices, but provides information about how to hedge out the default risk. The shortcoming of the structural approach is that it underestimates credit risk in the sense that the corporate bond is cheaper than the default free bond even if the firm value is larger than the default point. Subsequent to the work of \cite{M}, there have been many extensions such as \cite{BC}, \cite{LS}, etc. The second group of credit models, known as reduced form models, are more recent. These models, including the \cite{AD}, \cite{JT}, \cite{jow} and \cite{DS} models, do not look inside the firm. Instead, they assume that default occurs without warning at an exogenous default rate, or intensity. So, the default event would be unexpected. The dynamics of the intensity are specified under the pricing probability. Instead of asking why the firm defaults, the intensity model is calibrated using the market data such that the credit rating, the stock prices, risk free rate and etc. As the market data come not only from credit risk but also from other factors, thus the reduced form model overrates the credit risk. This is the shortcoming of the reduced approach.
One of the recent trends is to combine both approaches, see for example \cite{CE}, \cite{R}. In \cite{CE} a PDE method is used, providing a semi-analytical pricing formula of defaultable bond combining the two approaches when the short rate follows CIR model and the default intensity is linearly dependent on the short rate. 
The case with exogenous default recovery and no correlation between the firm value and the short rate is studied. It is assumed that the expected and unexpected default recovery rates are the same. 
\cite{R}, using PDE method, provided an analytical pricing formula of defaultable bond with both approaches in the case that the default intensity and the short rate are constant and the default recovery is endogenous. He also provided an analytical pricing formula when the short rate is uncorrelated with the firm value and the default recovery is exogenous. He didn't assumed that the unexpected and expected default recovery rates are the same.

However, a market with stochastic default intensity is incomplete in that the intensity is a source of uncertainty that is not traded. The incompleteness of the market usually gives rise to infinitely many martingales measures, each of which produces a no-arbitrage pricing. So, it is not clear which one to use in the pricing of the defaultable bond. The superreplication price (upper hedging price) is the minimal initial wealth needed for hedging the claim without risk. If the bank decides to charge a superreplication price for selling a bond so that it can trade to eliminate all risks, the price is usually high and unrealistic. There are three major approaches that have been developped in searching for solutions of pricing and hedging in incomplete markets. 

One is to pick a specific martingale measure for pricing according to some optimal criterion. For example on can cite the minimal relative entropy martingale measure in \cite{ave} or the q-optimal measure in \cite{hob}. The drawback is this approach sometimes provides a result that is not very reasonable financially. Another approach is to identify the martingale measure consistent with the market price. The problem is it does not provide a hedging strategy at all. Instead of choosing an equivalent martingale measure for valuation, a dynamic utility-based valuation theory has been developed producing the so-called Indifference price.

Utility indifference pricing was first introduced by \cite{HN}. It is an alternative where the price is uniquely determined at the cost of depending on the preferences of the pricer. The writer utility indifference price is the value of the initial payment that makes the seller indifferent to whether to sell the contract or not. From practical point of view utility indifference pricing has at least two advantages: 1) it does not refer to the market portfolio; 2) it generates an optimal hedging strategy in the sense that the resulting utility is equal to that of an optimal pure investment. 
The mathematical structure of utility indifference pricing has been well characterized by numerous researchers, cf. \cite{REK}, \cite{Delb}, \cite{Bech}, \cite{MUZA}, \cite{MASC}, and references therein.

\cite{BJ}, \cite{CH}, \cite{SZ}, \cite{SO} and \cite{leu} apply the utility-based pricing to credit risk. \cite{BJ} discussed the utility-indifference price of defaultable claims within the reduced approach by a backward stochastic differential equation. They studied a particular indifference price, based on the quadratic criterion and solved the problem by the duality approach for exponential utilities. \cite{CH} study the utility based prices of Event Sensitive Contingent (\emph{ESCC}) claims under two scenarios of resolution of uncertainty for event risk: when the event is continuously monitored or when it is revealed only at the payment date. In both cases, they transformed the incomplete market optimal portfolio choice problem of an agent endowed with an \emph{ESCC} into a complete market problem with a state and possibly path dependent utility function. They also obtain an explicit representation of the utility based prices under both information resolution scenarios in the case of negative exponential utility function. \cite{SZ} apply the utility indifference valuation in intensity-based models of default risk where the default time is the first jump of a time-changed Poisson process. They derive the Hamilton-Jacobi-Bellman (HJB) equations and analyze resulting yield spreads for single-name defaultable bonds, and a simple representative two-name credit derivative when the default intensity is constant. \cite{SO} discussed the utility-based pricing of defaultable bonds where their recovery values are unpredictable. He derived a partial integro-differential equations that the utility-based bond price solves. He also extracted credit risk premium from the yield spread of defaultable bonds and classified them to default-timing risk, recovery risk and spread risk. \cite{leu} apply the technology of utility-indifference valuation for defaultable bonds in a structural model for Black-Cox-type. They derive the HJB equations and simplify them to the linear (Feynman-Kac) differential equations. Finally, they find that the utility valuation has a significant impact on the bond prices and yield spreads.

The indifference valuation should be attractive to participants working in the OTC market. It is a direct way for them to quantify the default risks they face in a portfolio of complex instruments, when calibration data is scarce. Motivated by theses issues, our paper is similar to \cite{SZ} and \cite{leu} but here the default time $\tau$ is the first jump time of a Cox process with a random intensity $\lambda \geq 0$, which is correlated with the firm's stock price $S$. However, we consider the full filtration $\mathbb{F}=(\mathcal{F}_{t})_{t\geq 0}$ to make the default time $\tau$ a $\mathbb{F}$-stopping time. We also keep the same notation as \cite{SZ} and \cite{leu} to facilitate the comparison of the results. Assuming the intensity follows a \cite{CIR} (CIR) model, a finite difference method is used to solve the nonlinear reaction-diffusion equations satisfied by the value functions. The analysis of the bid and ask spread curves clearly shows the nonzero short-term yield-spread reflecting that the default risk comes as a surprise. This effect is not significative in \cite{leu} even if their results enhance short term yield spreads compared to the standard Black-Cox valuation. An interesting question is now, how the yield-spread behaves depending on different factors. It is quite obvious, that when the default intensity increases, the spread curves will be high. But what kind of behavior for the spreads would be expected with varying maturity to the exposure, given a default intensity? Some empirical studies have focused on a split behavior of the credit spreads according to the credit quality of the firm; see \cite{fons} and \cite{jow}. \cite{fons} using the rating information produced by Moody's, discovers an almost strictly upward slope for the spread curves for bonds in investment-grade classes. The so-called humped-shaped behavior can be observed for the rating BB and a downward slope has been calculated for the rating class B. \cite{jow} found for the issues rated AAA, AA or A a strictly upward sloping credit spread curve. For the issues rated BBB, BB and B a humped shape curve is obtained, while for the rating class CCC a downward sloping curve is observed. It is important to notify that in the precedent studies of split behavior, markets are assumed to be complete and free of arbitrage opportunities.

We apply the model to analyze the shape of the spread curve in the incomplete market according to the level of the default intensity. We found for the bid spread curve, a upward sloping curve for $0<\lambda\leq 0.06$, a $S$ shape curve for $0.06<\lambda\leq 0.15$ and a downward sloping curve for $0.15 < \lambda \leq 1$. For the ask spread curves, a upward sloping curve is observed for $0<\lambda\leq 0.25$ and a humped sloping curve for $0.25<\lambda\leq 1$. We also observed the positive relationship between the bid spread and the risk aversion parameter and a negative relationship between the ask spread and the risk aversion whatever the maturity. The impact of the correlation coefficient on the spread curves is analyzed. It shows the correlation coefficient increases the bid and ask spread for longer maturity. Finally we recover the classical structural models by proving they are the limit cases of our model.


The remainder of this paper is organized as follows. Section \ref{utility max} provides a brief introduction to utility maximization and indifference pricing. Section \ref{indifference} examines the indifference pricing of defaultable. Using the exponential utility function, the Hamilton-Jacobi-Bellman (\emph{HJB}) equations are presented and simplified to some reaction-diffusion equations. The dual optimization problem is revised and then some numerical results are presented.

\newpage

\section{Utility Maximization and Indifference Pricing}
\label{utility max}
\subsection{Utility Maximization}
 Utility maximization or the maximization of expected utility is a very basic topic in economics. A rigorous treatment of this subject within the framework of mathematical finance was initiated by the two seminal papers by \cite{ME} and \cite{SA}. Since then the problem of utility maximization has found its way into numerous textbooks : from the more economic point of view, e.g., \cite{Huli}, as well as from the more mathematical side, e.g., \cite{Kash} or \cite{Fsh}.
 
  Let us start by giving an introduction to the kind of questions we are going to treat in the following and describing the setting in which this will be done. We assume that there is an economic market described by a probability space $(\Omega,(\mathcal{F}_{t})_{0\leq t \leq T}, \mathbb{P})$. $\Omega$ denotes the possible states of the world, $\mathcal{F}_{t}$ the filtration which contains all information up time $t$, and $\mathbb{P}$ is the probability measure describing the likelihood of certain events. In the market there are a certain number of assets $(S_{t}^{i})_{0\leq t\leq T}$. We take one asset as numeraire, meaning that we assume that this special asset, called the bond or cash account, is constant equal to 1 and that the other assets are denoted in units of the bond. The other assets, in the following called stocks, are assumed to be risky and are described by an adapted, i.e., $\mathcal{F}_{t}$-measurable, stochastic processes. The market we consider is furthermore assumed to be frictionless. This means that the agents who trade in the market, i.e., buy and sell assets, do not face any transaction costs. Thus the gain or loss of an agent engaged in the market following a trading strategy $\theta$ is given by the stochastic integral $\int \theta dS$, also meaning that the trading strategy is self-financing, there is no money exogenously added or withdrawn. The agents face no constraints like short-selling constraints prohibiting them from holding a negative amount of a certain stock. In addition we always assume that the market is free of arbitrage, basically meaning that it is not possible to make money out of nothing.
  
  We now consider a special agent who has sold at time $0$ a contingent claim $C$ that matures at time $T$. Such contingent claim is modelled as an $\mathcal{F}_{T}$-measurable random variable. The agent will have to pay back the random amount of $C$ units of the bond at final time $T$. Let us also assume that his current wealth level is $c$. Thus the agent, if he chooses the trading strategy $\theta$, at final time T will hold
  
  $$Z(\theta)(\omega)=c+\int^{T}_{0}\theta(\omega)dS(\omega)-C(\omega)$$
  We call the random variable $Z(\theta)$ the outcome of the trading strategy $\theta$.
  
  Now, the question is how that the agent choose the trading strategy? We assume that his preferences have an expected utility representation, following the theory introduced by \cite{Nemo}. This means that he tries to maximize his expected utility over all
admissible trading strategies $\theta$:

$$E(U(Z(\theta)))\rightarrow \text{max}_{\theta}$$
  The \cite{Nemo} utility function $U(x)$ is assumed to be strictly increasing corresponding that investors prefer more wealth to less and strictly concave. The concavity reflects the issue of risk aversion: people are assumed to dislike risk and thus prefer for example to get 100 currency unit for sure compared to having a 50-50 chance of getting 200 currency unit or nothing. The concavity implies :
  
  $$\lambda U(x)+(1-\lambda)U(y)<U(\lambda x +(1-\lambda)y),\,\, \forall \lambda \in (0,1).$$
  In our example $\frac{1}{2}U(200)+\frac{1}{2}U(0)<U(100)$.\\
  Let us introduce the coefficient of absolute risk aversion defined by
  
   $$R_{a}(x)=-\frac{U''(x)}{U'(x)}.$$
   Since $U(x)$ is assumed to be strictly increasing, which means (provided $U(x)$ is sufficiently differentiable, which we assume) $U'(x)>0$, and strictly concave, i.e., $U''(x)<0$, the risk aversion $R_{a}(x)$ is greater than 0. The measure $R_{a}(x)$ of risk aversion was introduced by \cite{Arrow} and \cite{Pratt}. As indicated above it is very reasonable to use the curvature i.e. the degree of concavity $U''(x)$ as a measure for the risk aversion. The reason why $U''(x)$ is divided by $U'(x)$, which corresponds to a normalization, is the fact that \cite{Nemo} utility functions are only unique up to positive linear transformations. That is $U(x)$ and $\bar{U}=aU(x)+b$ for $a,\, b \in \mathbb{R},\,\,a>0$ correspond to the same preference relation of the agent. Dividing $U''(x)$ by the derivative $U'(x)$ respectively $\bar{U}'(x)$ ensures the same risk aversion coefficient. There exist many utility functions and the most useful in utility maximization are power utility functions $U(x)=\frac{x^{1-R}}{1-R}$; $R \neq 1$, $R >0$ and exponential utility $U(x)=-e^{-\gamma x}$; $\gamma >0$. The description of both of these functions is given by describing the \emph{HARA} and \emph{Non-HARA} utilities.\\
 
 A utility function is of the HARA class if $R_{a}(x)=\frac{1}{A+Bx}$; $x \in I_{D}$ where $I_{D}$ is the interval on which $U$ is defined and $B$ is a non-negative constant.
 
\begin{itemize}
\item   If $B >0$ and $B \neq 1$, this leads under some assumptions to the standard power utility function $U(x)=\frac{x^{1-R}}{1-R}$. The power utility has constant relative risk aversion (CRRA) of $R$, where the relative risk aversion $R_{r}(x)=xR_{a}(x)$. The power utility has fundamental assumption that it requires wealth to be non-negative.
 \item If $B=1$, we have under some assumptions the logarithmic utility function $U(x)=\text{ln}(x)$ which also requires a non-negative wealth.
 \item If $B=0$, this leads to the exponential utility function $U(x)=-e^{-\gamma x}$ which has a constant absolute risk aversion (CARA) of $\gamma$. Bigger $\gamma$ corresponds to higher degree of risk aversion. In particular, $\gamma\rightarrow \infty$ indicates absolute risk aversion while $\gamma\rightarrow 0$ corresponds to risk neutrality. In contrast to the power utility function, the exponential utility function allows for negative wealth.
\end{itemize}
 
The Non-HARA class regroups all utility functions which do not fit the HARA class. An utility function which belongs to the Non-HARA is the quadratic utility function $U(x)=x-bx^{2}$ where $b>0$. This utility function decreases over part of the range, violating the assumption that investors desire more wealth, but has excellent tractability properties; see \cite{Vicky}.
    
 \subsection{Utility Indifference Prices}
 \label{utility}
The utility indifference buy (or bid) price $p^{b}$ is the price at which the investor is indifferent, in terms of maximum expected utility, between paying nothing and not having the claim $C_{T}$ and paying $p^{b}$ now to receive $k >0$ units of the claim $C_{T}$ at time $T$. Assume that the investor initially has wealth $x$ and zero endowment of the claim. Mathematically, this expressed as
\begin{equation}
\text{sup}_{X_{T} \in A(x)}E[U(X_{T})]=\text{sup}_{X_{T} \in A(x-p^{b}(k))}E[U(X_{T}+kC_{T})],
\label{prob}
\end{equation}
where $A(x)$ is the set of all wealth $X_{T}$ which can be generated from initial fortune $x$.
Let $M(x)= \text{sup}_{X_{T} \in A(x)}E[U(X_{T})]$ and $H^{b}(x,k)=\text{sup}_{X_{T} \in A(x)}E[U(X_{T}+kC_{T})]$; the utility indifference buy price $p^{b}$ is the solution to 
\begin{equation}
M(x)=H^{b}(x-p^{b}(k),k),
\label{m=h_bid}
\end{equation}
that is the investor is willing to pay at most the amount $p^{b}(k)$ today for $k$ units of the claim $C_{T}$ at time $T$. Similarly the utility indifference sell (or ask) price $p^{s}(k)$ is the smallest amount the investor is willing to accept in order to sell $k$ units of $C_{T}$. The utility indifference sell price $p^{s}(k)$ solves the problem
\begin{equation}
M(x)=H^{s}(x+p^{s}(k),k),
\label{m=h_ask}
\end{equation}
where $H^{s}(x,k)=\text{sup}_{X_{T} \in A(x)}E[U(X_{T}-kC_{T})]$.\\
 
From equations (\ref{m=h_bid}) and (\ref{m=h_ask}), to compute the utility indifference price of a claim, two stochastic control problems must be solved. The first one is the optimal investment problem when the investor has a zero position in the claim; this is equivalent to the \cite{merton} optimal investment problem  and the value function is denoted by $M(x)$. The second one is the optimal investment problem when the investor has bought or sold $k$ units of the claim and is denoted by $H(x,k)$.\\

Utility indifference prices have some appealing properties. Theses properties are given in the following
\begin{theorem}
\begin{enumerate}
\item  \emph{Non-linear pricing.}
Firstly, in contrast to the arbitrage-free prices, utility indifference prices are non-linear in the number of claims, $k$. This means that the buyer is not willing to pay twice as much for twice as many claims, but requires a reduction in this price to take an additional risk. Alternatively, a seller requires more than twice the price for taking on twice the risk. This property can be seen by the value function $H$ since the utitlity function $U$ is concave.
\item \emph{Recovery of complete market price.}
\label{rec}
If the market is complete or the claim is perfectly replicable, the utility indifference price is equivalent to the complete market price for $k$ units.
\item  \emph{Monotonicity.}
\label{rec1}
Let $p^{1}$ and $p^{2}$ be the utility indifference prices for one unit of payoff $C_{T}^{1}$ and $C_{T}^{2}$ respectively and $C_{T}^{1} \leq C_{T}^{2}$, then $p^{1} \leq p^{2}$.
\item \emph{Concavity.}
\label{rec2}
Let $p_{\lambda}$ be the utility indifference bid price for the claim $\lambda C_{T}^{1} +(1-\lambda)C_{T}^{2}$ where  $\lambda \in [0,1]$, then 
$$p_{\lambda} \geq \lambda p^{1} +(1-\lambda)p^{2}$$
\end{enumerate}
\end{theorem}
We briefly recall the proof of theses results. (See also \cite{Vicky}) 
\begin{proof} (\ref{rec})
 Let $C_{T}$ the contingent claim which is replicable, $p^{c}$ the arbitrage-free price of the claim and $p^{b}(k)$ the buyer indifference price of $k$ units of the contingent claim. We denote by $R_{T}$ the time $T$ value of unit of currency invested at time $0$. Since $X_{T} \in A(x)$, $X_{T}=x R_{T}+\tilde{X_{T}}$ for some $\tilde{X_{T}} \in A(0)$, where $A(0)$ is the set of claims which can be replicated by zero initial wealth. Since $C_{T}$ is replicable from an initial fortune $p^{c}$, we have $C_{T}=p^{c} R_{T}+\tilde{X_{T}}^{c}$ where $\tilde{X_{T}}^{c} \in A(0)$. Then for $X_{T} \in A(x)$, $X_{T}+k C_{T}=(x+k p^{c}) R_{T}+\tilde{X_{T}}+k \tilde{X_{T}}^{c}=(x+k p^{c}) R_{T}+\tilde{X_{T}}'$ where $\tilde{X_{T}}'=\tilde{X_{T}}+k \tilde{X_{T}}^{c} \in A(0)$. Thus $X_{T}+k C_{T} \in A(x+k p^{c})$. But $H^{b}(x,k)=\text{sup}_{X_{T} \in A(x)}E[U(X_{T}+kC_{T})]=\text{sup}_{X_{T} \in A(x+k p^{c})}E[U(X_{T})]=M(x+k p^{c})$ and thus $p^{b}(k)=kp^{c}$. It means that the buyer indifference price is $k$ units times the arbitrage-free price. 
\end{proof}
 
\begin{proof} (\ref{rec1}) Let assume that $C_{T}^{1} \leq C_{T}^{2}$, then  $H^{b}(x,1,C_{T}^{1})\leq H^{b}(x,1,C_{T}^{2})$, with $H^{b}(x,1,C_{T}^{1})$ and $H^{b}(x,1,C_{T}^{2})$ are buyer's value functions for one unit of the claims $C_{T}^{1}$ and $C_{T}^{2}$ respectively. As $H^{b}(x-p^{1},1,C_{T}^{1})=H^{b}(x-p^{2},1,C_{T}^{2})$ on deduce  $M(x+p^{1})\leq M(x+p^{2})$ and thus $p^{1}\leq p^{2}$.
\end{proof}
\begin{proof}(\ref{rec2}) 
Let $X_{T}^{i}$ be the optimal target wealth for an individual with initial wealth $x-p^{i}$ due to receive the claim $C_{T}^{i}$. Then,
$M(x)=H^{b}(x-p^{i},1,C_{T}^{i})=\text{sup}_{X_{T} \in A(x-p^{i})}E[U(X_{T}+C_{T}^{i})]=E[U(X_{T}^{i}+C_{T}^{i})]$. Define $\bar{X_{T}}=\lambda X_{T}^{1}+(1-\lambda) X_{T}^{2}$, we have $\bar{X_{T}} \in \bar{A}=A(x-\lambda p^{1} -(1-\lambda)p^{2})$. So,\\
$H^{b}(x-\lambda p^{1} -(1-\lambda)p^{2},1,\lambda C_{T}^{1} +(1-\lambda)C_{T}^{2})$
\begin{eqnarray*}
&=&\text{sup}_{X_{T} \in \bar{A}}E[U(X_{T}+\lambda C_{T}^{1} +(1-\lambda)C_{T}^{2})]\\
&\geq& E[U(\bar{X_{T}}+\lambda C_{T}^{1} +(1-\lambda)C_{T}^{2})]\\
&=&E[U(\lambda (X_{T}^{1}+C_{T}^{1})+(1-\lambda)(X_{T}^{2}+C_{T}^{2})]\\
&\geq& \lambda E[U(X_{T}^{1}+C_{T}^{1})]+(1-\lambda) E[U(X_{T}^{2}+C_{T}^{2})]\\
&=&M(x)\\
&=& H^{b}(x-p_{\lambda},1,\lambda C_{T}^{1} +(1-\lambda)C_{T}^{2})
\end{eqnarray*}
Thus
$H^{b}(x-\lambda p^{1} -(1-\lambda)p^{2},1,\lambda C_{T}^{1} +(1-\lambda)C_{T}^{2})\geq H^{b}(x-p_{\lambda},1,\lambda C_{T}^{1} +(1-\lambda)C_{T}^{2})$ and therefore $p_{\lambda} \geq \lambda p^{1} +(1-\lambda)p^{2}$
\end{proof}

If we consider ask price rather bid price then $p_{\lambda}$ is convex rather than concave.

These value functions are the solutions of the Hamilton-Jacobi-Bellman (HJB) equations. After solving theses equations, we can find the utility indifference bid price and ask price for the claim. In the next chapter, we will derive the utility indifference for defaultable bonds in the case of stochastic intensity.

\section{Indifference Pricing For Defaultable Bonds}
\label{indifference}

\subsection{Maximal Expected Utility Problem}
\label{max_exp}
Let us start with a defaultable bond of a firm with expiration date $T < \infty$. Unlike in a traditional structural approach, default time $\tau$ is the first jump of a Cox process with a random intensity $\lambda \geq 0$, which is correlated with the firm's stock price $S$. The price of the asset $S$ is described by a geometric Brownian motion. The intensity process is $\lambda (Y_{t})$ where $\lambda(.)$ is a non-negative, locally Lipschiptz, smooth and bounded function. The dynamics of $S$ and $Y$ are
\begin{eqnarray}\left\{\begin{array}{lcl}
dS_{t}&=& \mu S_{t} dt+\sigma S_{t} dW_{t}^{1}, \,\,S_{0}=S>0,\\
dY_{t}&=&b(Y_{t})dt + a(Y_{t})\left(\rho dW_{t}^{1}+\rho'dW_{t}^{2}\right), \,\,Y_{0}=\lambda,
\label{asset_price_bm}
\end{array}\right.
\end{eqnarray}
with $\rho'=\sqrt{1-\rho^{2}}$.
The processes $W^{1}$ and $W^{2}$ are two independent Brownian motions defined on a probability space $(\Omega,\mathcal{F},P)$ and $\mathcal{F}_{t}$ is the large filtration enough to support a counting process $N_{t}=1_{\left\{\tau<t\right\}}$ and stochastic processes (($S_{u}$, $Y_{u})$; $0 \leq u \leq t)$ which are independent of $N$. Hence we may write $\mathcal{F}_{t}=\mathcal{G}_{t}\vee \mathcal{H}_{t}$, where $\mathcal{G}_{t}=\sigma\left\{(S_{u}, Y_{u}): 0 \leq u \leq t \right\}$ and $\mathcal{H}_{t}=\sigma\left\{N_{u}:0 \leq u \leq t\right\}$. The background filtration $\mathcal{G}_{t}$ contains all economic informations except for the default times available to investors up to time $t$. However, the default time $\tau$ is not a stopping time with respect to $\mathcal{G}_{t}$ but it is under $\mathcal{F}_{t}$. In addition, if we condition on the filtration $\mathcal{G}_{t}$, $N_{t}$ is an inhomogeneous Poisson process with the time-varying intensity $\lambda$. 
The parameter $\rho \in (-1,1)$ measures the instantaneous correlation between shocks to the stock price $S$ and shocks to the intensity-driving process $Y$. We also assume that there exists a unit exponential random variable $E_{1}$, independent of the Brownian motions. The default time $\tau$ of the firm is defined by 
$$\tau=\text{inf} \left\{s \geq 0:\int^{s}_{0} \lambda_{u}du\geq E_{1} \right\}.$$
We make the following assumption:
\begin{Assumption}
 The coefficients $a(.)$ and $b(.)$ are such that (\ref{asset_price_bm}) has a unique strong solution $Y_{t}$ which lies in $\mathbb{R}_{+}$ for all $t \in [0,T]$, P-a.s.
\end{Assumption}
Observe that Assumption $1$ is weaker than the usual global Lipschitz continuity and growth conditions that are sufficient for the existence of a strong solution $Y_{t}$. This avoids ruling out from the stochastic intensity model such that the CIR model, for which $Y_{t}$ is square root diffusion that does not satisfy the global Lipschipz condition.\\
Assume that a bank account is also traded in the market and the risk free rate interest rate is $r$. The investor's control process is $\pi_{t}$, the amount held in the stock at time $t$, until $\tau \wedge T$. The control process $\pi$ is admissible i.e. it is $\mathcal{F}_{t}$ measurable and satisfies the integrability contraint $E\{\int_{0}^{T}\pi_{s}^{2}ds\}<\infty$. The set of all admissible strategies is denoted by $A$. In $t <\tau \wedge T$, the investor's wealth process $X_{t}$ follows 
\begin{eqnarray}
dX_{t}&=&\pi_{t}\frac{dS_{t}}{S_{t}}+r(X_{t}-\pi_{t})dt \\
&=&(rX_{t}+\pi_{t}(\mu-r))dt+\sigma \pi_{t}dW_{t}^{1}.  
\label{x_equation_1}
\end{eqnarray}
If the default event occurs before $T$, the investor has to liquidate holdings on the stock and deposit in the bank account. For simplicity, we assume he receives full pre-default market value on his stock holdings on liquidation. 
Therefore, given that $\tau < T$, for $\tau \leq t \leq T$, we have $X_{t}=X_{\tau}e^{r(t-\tau)}.$

\begin{figure}[H]
\begin{center}
\rotatebox{0}{\includegraphics[width=3in]{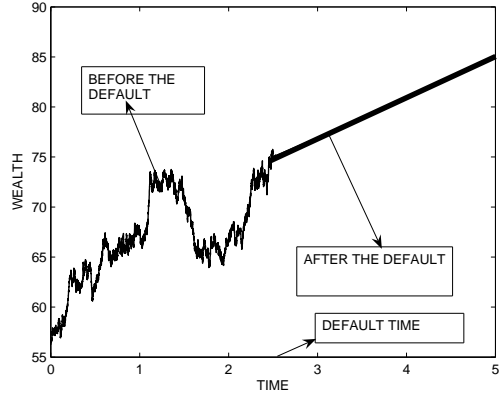}}
\caption{Evolution of Wealth before and after the default event}
\label{fig1}
\end{center}
\end{figure}

Using the exponential utility of discounted wealth, we are first interested in the optimal investment problem up to time $T$ of the investor who does not hold the defaultable bond. At time zero, the maximum expected utility of discount payoff takes the form:
\begin{equation}
\text{sup}_{\pi \in A}E\left(-e^{-\gamma (e^{-rT}X_{T})}1_{\{\tau >T\}}+(-e^{-\gamma(e^{-r \tau}X_{\tau})})1_{\{\tau \leq T\}} \right)
\label{merton}
\end{equation}
To simplify the discount variable, let us apply the It\^o Lemma to $e^{-rt}X_{t}$ 
$$d(e^{-rt}X_{t})=e^{-rt}(\pi_{t}(\mu-r)dt+\sigma \pi_{t}dW_{t}^{1}),$$
and switching the discount variable $X_{t}\rightarrow e^{-rt}X_{t}$, the excess return $\mu \rightarrow \mu-r$ and $\pi_{t} \rightarrow \pi_{t}e^{-rt}$, we get 
\begin{equation}
dX_{t}=\pi_{t}\left(\mu dt+ \sigma dW_{t}^{1}\right).
\label{x_eq_2}
\end{equation}
Next, we define the stochastic control problem initiated at time $t\leq T$, and define the default time 
$$\tau_{t}=\text{inf} \left\{s \geq t:\int^{s}_{t} \lambda_{u}du\geq E_{1} \right\}.$$
The Merton value function becomes
\begin{equation}
M(t,x,y)=\text{sup}_{\pi_{t} \in A}E_{t}\left(-e^{-\gamma X_{T}}1_{\{\tau_{t} >T\}}+(-e^{-\gamma X_{\tau_{t}}})1_{\{\tau_{t} \leq T\}} \right),
\label{merton2}
\end{equation}
where $(x,y)=(X_{t},Y_{t})$, $E_{t}(\cdot)=E(\cdot \vert \mathcal{F}_{t})$.\\

We are now interested to the same problem of the investor who owns the defaultable bond of the firm, which pays 1 currency unit on date $T$ if the firm has survived till then. The bond's holder value function is:
\begin{equation}
H^{b}(t,x,y)=\text{sup}_{\pi_{t} \in A}E_{t}\left(-e^{-\gamma (X_{T}+c)}1_{\{\tau_{t} >T\}}+(-e^{-\gamma X_{\tau_{t}}})1_{\{\tau_{t} \leq T\}} \right),
\label{val_fct_invest}
\end{equation}
where $c=e^{-rT}$.
In the same way, the bond's seller value function is :
\begin{equation}
H^{s}(t,x,y)=\text{sup}_{\pi_{t} \in A}E_{t}\left(-e^{-\gamma (X_{T}-c)}1_{\{\tau_{t} >T\}}+(-e^{-\gamma X_{\tau_{t}}})1_{\{\tau_{t} \leq T\}} \right).
\label{val_fct_sell}
\end{equation}
\begin{remark}
\item The  value functions $M$, $H^{b}$, $H^{s}$ are concave, increasing in $x$ and uniformly bounded in $y$. See \cite{SZ}.
\end{remark}

In the next section, we derive the Hamilton-Jacobi-Bellman equations satisfied by these value functions.

\subsection{The Hamilton-Jacobi-Bellman Equations} 
\label{hjb_eq}

\begin{proposition}
Consider the Merton value function 
$$M(t,x,y)=\text{sup}_{\pi_{t} \in A}E_{t}\left(-e^{-\gamma X_{T}}1_{\{\tau_{t} >T\}}+(-e^{-\gamma X_{\tau_{t}}})1_{\{\tau_{t} \leq T\}} \right),$$
subject to the wealth and state variable contraints
$$dX_{t}=\pi_{t}\left(\mu dt+\sigma dW_{t}^{1}\right)$$
and
$$dY_{t}=b(Y_{t})dt + a(Y_{t})\left(\rho dW_{t}^{1}+\rho'dW_{t}^{2}\right)$$
with the coefficients $a(.)$ and $b(.)$  verifying the assumption $1$.\\
The Merton value function $M$ is the unique viscosity solution in the class of functions that are concave and increasing in $x$, and uniformly bounded in $y$ of the Hamilton-Jacobi-Bellman (HJB) equation
\begin{eqnarray}
\text{sup}_{\pi \in A}\left\{\frac{E_{t}\left[dM(t,x,y)\right]}{dt}\right\}=0
\end{eqnarray}
\end{proposition}

\begin{proof} Assuming that an optimal amount exists, we derive from 
$$\text{sup}_{\pi_{t} }E_{t}\left(-e^{-\gamma X_{T}}1_{\{\tau_{t} >T\}}+(-e^{-\gamma X_{\tau_{t}}})1_{\{\tau_{t} \leq T\}} \right)-M(t,x,y)=0$$
for some small $h>0$,
$$\text{sup}_{\pi_{t} }E_{t}\left[E_{t+h}\left(-e^{-\gamma X_{T}}1_{\{\tau_{t} >T\}}+(-e^{-\gamma X_{\tau_{t}}})1_{\{\tau_{t} \leq T\}}\right)\right]-M(t,x,y)=0.$$
The expression $E_{t+h}\left(-e^{-\gamma X_{T}}1_{\{\tau_{t} >T\}}+(-e^{-\gamma X_{\tau_{t}}})1_{\{\tau_{t} \leq T\}}\right)$ is nothing else than the expected utility for an investor starting with wealth $X_{t+h}$ at time $t+h$. Therefore, for any control $\pi_{s}$ with $s \geq t+h$,
$$E_{t+h}\left(-e^{-\gamma X_{T}}1_{\{\tau_{t} >T\}}+(-e^{-\gamma X_{\tau_{t}}})1_{\{\tau_{t} \leq T\}}\right)\leq M(t+h,X_{t+h},Y_{t+h}),$$
with equality holds for the optimal $\pi^{*}$. Hence,
$$\text{sup}_{\pi_{s},t\leq s\leq t+h }E_{t}\left[M\left(t+h,X_{t+h},Y_{t+h}\right) \right]-M(t,x,y)=0.$$
Assuming an optimal behavior from $t+h$ on $T$, the optimal $\pi^{*}$ has only to be determined until $t+h$, and not on the whole time horizon. Dividing by $h$ and applying the limit $h\rightarrow 0$, the above equation becomes
$$\text{sup}_{\pi_{t}}\left\{\text{lim}_{h\rightarrow 0}E_{t}\frac{1}{h}\left[M\left(t+h,X_{t+h},Y_{t+h}\right)-M(t,x,y)\right]\right\}=0$$
Using the theorem of bounded convergence, this equation becomes
$$\text{sup}_{\pi_{t}}\left\{E_{t}\text{lim}_{h\rightarrow 0} \frac{1}{h} \left[M\left(t+h,X_{t+h},Y_{t+h}\right)-M(t,x,y)\right]\right\}=0,$$
but $\text{lim}_{h\rightarrow 0}\frac{1}{h}\left[M\left(t+h,X_{t+h},Y_{t+h}\right)-M(t,x,y)\right]=\frac{d}{dt}M(t,x,y)$ and the value function $M$ verifies
$$\text{sup}_{\pi \in A}\left\{\frac{E_{t}\left[dM(t,x,y)\right]}{dt}\right\}=0$$
(see \cite{SEN} or \cite{MB} for more details).\\
Following the arguments of Theorem $4.1$ and $4.2$ in \cite{DZ}, we deduce that the value function $M$ is the unique viscosity solution of the $HJB$
\end{proof}

Let us now derive the explicit form of the HJB equations satisfied by the Merton function $M$.
We have
\begin{multline*}
E_{t}\left(dM(t,x,y)\right)=\\
\left(M_{t}+\pi \mu M_{x}+b(y)M_{y}+\frac{1}{2}\sigma^{2} \pi^{2} M_{xx}+\frac{1}{2} a(y)^{2} M_{yy}+\rho \sigma \pi a(y) M_{xy}\right)dt+\\
\lambda(y) \left(-e^{-\gamma x}-M\right)dt,
\end{multline*}
Dividing by $dt$ and applying the supremum, we obtain
\begin{eqnarray}
M_{t}+\mathcal{L}_{y}M+\text{sup}_{\pi}\left(\frac{1}{2}\sigma^{2}\pi^{2}M_{xx}+\pi\left(\rho \sigma a(y)M_{xy}+\mu M_{x}\right)\right)\nonumber \\
+\lambda(y)\left(-e^{-\gamma x}-M \right)=0
\end{eqnarray}
with $M(T,x,y)=-e^{-\gamma x}$ and $\mathcal{L}_{y}=\frac{1}{2}a(y)^{2}\frac{\partial^{2}}{\partial y^{2}}+b(y)\frac{\partial}{\partial y}$

Similarly, the value function $H^{b}$ of the investor, who owns the defaultable bond, satisfies the HJB equation is:
\begin{eqnarray}
H^{b}_{t}+\mathcal{L}_{y}H^{b}+\text{sup}_{\pi}\left(\frac{1}{2}\sigma^{2}\pi^{2}H^{b}_{xx}+\pi\left(\rho \sigma a(y)H^{b}_{xy}+\mu H^{b}_{x}\right)\right)\nonumber\\
+\lambda(y)\left(-e^{-\gamma x}-H^{b} \right)=0
\end{eqnarray}
with $H^{b}(T,x,y)=-e^{-\gamma(x+c)}$. 

The value function $H^{s}$ of the seller of the bond also satisfies :
\begin{eqnarray}
H^{s}_{t}+\mathcal{L}_{y}H^{s}+\text{sup}_{\pi}\left(\frac{1}{2}\sigma^{2}\pi^{2}H^{s}_{xx}+\pi\left(\rho \sigma a(y)H^{s}_{xy}+\mu H^{s}_{x}\right)\right)\nonumber\\
+\lambda(y)\left(-e^{-\gamma x}-H^{s} \right)=0,
\end{eqnarray}
 with $H^{s}(T,x,y)=-e^{-\gamma(x-c)}$.
 
Using (\ref{m=h_bid}) and (\ref{m=h_ask}), the buyer's indifference price $p^{b}_{0}(T)$ and the seller's indifference price $p^{s}_{0}(T)$ (at time $0$) of the defaultable bond with maturity $T$ are defined by 
\begin{eqnarray}
M(0,x,y)=H^{b}(0,x-p^{b}_{0},y) 
\end{eqnarray}
and 
\begin{eqnarray}
M(0,x,y)=H^{s}(0,x+p^{s}_{0},y).
\end{eqnarray}

\begin{remark}
\item It is well-known that the indifference price under exponential utility does not depend on the investor's initial wealth $x$, which is an attractive feature of using this utility function.
\end{remark}
In the next section we factorize the value functions $M$, $H^{b}$, $H^{s}$, that leads to some reaction-diffusion equations.

\subsection{Reaction-Diffusion Equations} 
\label{reac_diff}

In order to simplify the above HJB equations let us introduce the familiar distortion scaling:
$$M(t,x,y)=-e^{-\gamma x} u(t,y)^{\frac{1}{1-\rho^{2}}},$$  
$$H^{b}(t,x,y)=-e^{-\gamma(x+c)} w^{b}(t,y)^{\frac{1}{1-\rho^{2}}},$$ 
and 
$$H^{s}(T,x,y)=-e^{-\gamma(x-c)} w^{s}(t,y)^{\frac{1}{1-\rho^{2}}},$$
with $u,w^{b},w^{s}:[0,T]\times \mathbb{R}\rightarrow \mathbb{R^{+}}$. For the proof of the factorization, see \cite{Z}. Plugging these expressions in the HJB equations, we get the following reaction-diffusion equations: 
\begin{eqnarray}
u_{t}+\mathcal{K}_{y}u-\left(1-\rho^{2}\right)\left(\frac{\mu^{2}}{2 \sigma^{2}}+\lambda(y)\right)u+\left(1-\rho^{2}\right)\lambda(y)u^{-\theta}=0,
\label{M-function}
\end{eqnarray}
with the final condition $u\left(T,y\right)=1$.
\begin{eqnarray}
w^{b}_{t}+\mathcal{K}_{y}w^{b}-\left(1-\rho^{2}\right)\left(\frac{\mu^{2}}{2 \sigma^{2}}+\lambda(y)\right)w^{b}+\left(1-\rho^{2}\right)e^{\gamma c}\lambda(y)(w^{b})^{-\theta}=0,
\label{M1-function}
\end{eqnarray}
with the final condition $w^{b}\left(T,y\right)=1$.
\begin{eqnarray}
w^{s}_{t}+\mathcal{K}_{y}w^{s}-\left(1-\rho^{2}\right)\left(\frac{\mu^{2}}{2 \sigma^{2}}+\lambda(y)\right)w^{s}+\left(1-\rho^{2}\right)e^{-\gamma c}\lambda(y)(w^{s})^{-\theta}=0
\label{M2-function}
\end{eqnarray}
with the final condition $w^{s}\left(T,y\right)=1$,
where 
$$\theta=\frac{\rho^{2}}{1-\rho^{2}},$$
and 
$$\mathcal{K}_{y}=\mathcal{L}_{y}-\frac{\rho \mu}{\sigma}a(y)\frac{\partial}{\partial y}.$$

With the simplified form of the value functions, the buyer indifference's price is given in the following way; from section \ref{utility},
$$M(0,x,y)=H^{b}(0,x-p^{b}_{0},y)$$
$$e^{-\gamma x} u(0,y)^{\frac{1}{1-\rho^{2}}}=e^{-\gamma(x-p_{0}^{b}+c)} w^{b}(0,y)^{\frac{1}{1-\rho^{2}}},$$
this yields to
\begin{eqnarray}
p_{0}^{b}(T)=e^{-rT}-\frac{1}{\gamma (1-\rho^{2})}\text{log}\left(\frac{w^{b}(0,y)}{u(0,y)}\right).
\end{eqnarray}
It is shown in  \cite{SZ} that $w^{s}(t,y)<u(t,y)\leq w^{b}(t,y)$ for $(t,y) \in \left[0,T\right] \times \mathbb{R}$, so $p_{0}^{b}(T) \leq e^{-rT}$.
The yield spread of the bond is defined by
\begin{eqnarray}
y_{0}^{b}(T)=-\frac{1}{T}\text{log}(p_{0}^{b}(T))-r,
\end{eqnarray}
which is non-negative for all $T$.
 
In the same way from section \ref{hjb_eq}, the seller indifference's price is 
\begin{eqnarray}
p_{0}^{s}(T)=e^{-rT}-\frac{1}{\gamma (1-\rho^{2})}\text{log}\left(\frac{u(0,y)}{w^{s}(0,y)}\right),
\end{eqnarray}
and the seller's yield spread is non-negative for all $T$.\\
  The indifference price in this model can be characterized as the solution of the non-linear reaction-diffusion equations related to the functions $u$, $w^{b}$ and $w^{s}$. We present in the next section the corresponding dual problems which is an alternative approach to evaluate the indifference price.
 
 \subsection{Dual Problem: Relative entropy Minimization}
 
 We recall the primal optimal investment problem for the buyer at time zero,
 
 \begin{equation*}
\mathcal{P}=\text{sup}_{\pi \in A}E\left(-e^{-\gamma (X_{T}+c)}1_{\{\tau_ >T\}}+(-e^{-\gamma X_{\tau}})1_{\{\tau \leq T\}} \right).
\end{equation*}
Under the framework that the underlying processes are locally bounded semimartingales and the utility function is exponential, \cite{Delb} proved the duality relation 

$$\mathcal{P}=-e^{-\gamma x-\gamma \mathcal{D}}$$

where $x$ is the initial wealth and $\mathcal{D}$ is the value of the dual optimization problem

\begin{equation}
\mathcal{D}=\text{inf}_{Q \in \mathbb{P}_{f}}\left(\mathbb{E}^{Q}\left\{c\right\}+\frac{1}{\gamma}\mathcal{H}(Q \vert P)\right)
\label{val_dual}
\end{equation}

where $\mathcal{H}(Q \vert P)$ is the relative entropy between $Q$ and $P$, namely,

\begin{eqnarray}
\mathcal{H}(Q \vert P)=\left\{ \begin{array}{lcl}\mathbb{E}\left\{\frac{dQ}{dP}\text{log}\left(\frac{dQ}{dP}\right)\right\}, Q<<P,\\ \infty,\,\,\,\,\,\,\,\,\,\,\,\,\, \text{otherwise}
\end{array}\right.
\end{eqnarray}
where $Q<<P$ denotes $Q$ is absolutely continuous with respect $P$.
In (\ref{val_dual}), $\mathbb{P}_{f}$ denotes the set of absolutely continuous local martingale measures with finite relative entropy with respect to $P$. 

\begin{Assumption}
There is an equivalent local martingale measure with finite relative entropy. Thus,
\begin{eqnarray}
\mathbb{P}_{f}(P)\cap \mathbb{P}_{e}\neq \phi
\end{eqnarray}

where $\mathbb{P}_{e}$ is the set of equivalent local martingale measures.
\label{amin}
\end{Assumption}

Using the exponential utility for discounted wealth, it is easy to show that 

\begin{eqnarray}
P_{0}^{b}(T)=\frac{1}{\gamma}\text{log}\left(\frac{M(0,x,y)}{H^{b}(0,x,y)}\right)
\end{eqnarray}
and substituting the specific expressions of the duals for the buyer and the Merton problems, we can write
\begin{eqnarray}
P_{0}^{b}(T)=\mathcal{D}-\frac{1}{\gamma}\text{inf}_{Q \in \mathbb{P}_{f}}\mathcal{H}(Q \vert P)
\label{inmin}
\end{eqnarray}


 The entropy term in (\ref{inmin}) can be combined into one entropy term with a different prior measure, the minimal entropy martingale measure, which is the measure minimizing the relative entropy in $\mathbb{P}_{f}$. Let 

 \begin{eqnarray}
Q^{0}=\text{arg min}_{Q \in \mathbb{P}_{f}}\mathcal{H}(Q \vert P)
\end{eqnarray}

\begin{theorem}
(\cite{Fri} and \cite{Delb}) Under assumption (\ref{amin}), $Q^{0}$ exists, is unique, is in $\mathbb{P}_{f}(P)\cap \mathbb{P}_{e}$ and its density has the form

\begin{eqnarray}
\frac{dQ^{0}}{dP}=\alpha_{0}e^{-\gamma X_{T}^{0}}
\end{eqnarray}

where $X_{T}^{0}$ is the optimal terminal wealth associated with the solution of the Merton optimization problem (\ref{merton2}) and $\log\alpha_{0}=\mathcal{H}(Q^{0} \vert P)<\infty$
\end{theorem}

Under the assumption that $\frac{dQ^{0}}{dP} \in L^{2}(P)$, we can apply Proposition $2.2$ of \cite{sij} and the bid price is
\begin{eqnarray}
P_{0}^{b}(T)=\text{inf}_{Q \in \mathbb{P}_{f}(Q^{0})}\left(\mathbb{E}^{Q}\left\{c\right\}+\frac{1}{\gamma}\mathcal{H}(Q \vert Q^{0})\right).
\end{eqnarray}

Next, we define the Radon-Nikodym derivative between $Q^{0}$ and $P$ for our model. This approach is taken in \cite{sij} and \cite{BJ}. The so-called minimal martingale measure, $P^{0}$, is defined by the following Girsanov transformation. 

$$\frac{dP^{0}}{dP}=\text{exp}(-\frac{\mu}{\sigma}W_{T}^{1}-\frac{\mu^{2}}{2\sigma^{2}}T)$$

The dynamic of $(S,Y)$ under $P^{0}$ are:
\begin{eqnarray}\left\{\begin{array}{lcl}
dS_{t}&=& r S_{t} dt+\sigma S_{t} dW_{t}^{P_{0}(1)}\\
dY_{t}&=&\left(b(Y_{t})-\frac{\rho \mu}{\sigma}a(Y_{t}) \right)dt + a(Y_{t})\left(\rho dW_{t}^{P_{0}(1)}+\rho'dW_{t}^{P_{0}(2)}\right),
\label{dist_Y}
\end{array}\right.
\end{eqnarray}

where \\

$W_{t}^{P_{0}(1)}=W_{t}^{1}+\frac{\mu}{\sigma}t$ and $W_{t}^{P_{0}(2)}=W_{t}^{2}$\\

$P^{0}$ has finite relative entropy and is equivalent to $P$. Therefore, $Q^{0} \in \mathbb{P}_{f}(P)\cap \mathbb{P}_{e}$ and is unique. The parameters $\mu$ and $\sigma$ are independent of $Y$ and by \cite{sij}, the minimal entropy martingale measure coincides with the minimal martingale measure.

\subsection{Comparison with the classical reduced model}
 In this section, we compare the indifference price for both buyer and seller with the classical reduced price. In fact, the latter is just the limit case of our model. Let us recall the classical results of the intensity model. See \cite{bie}.\\
 Let $Q$ be a risk neutral probability and $\lambda_{t}^{Q}$ the corresponding risk-neutral intensity. The price at time $t=0$ of the defaultable bond denoted $P^{ar}_{0}(T)$ is thus equal to the expectation under $Q$ of the discount pay-off,
  
  \begin{eqnarray*}
  P^{ar}_{0}(T)=E^{Q}(e^{-rT}1_{\left\{\tau>T\right\}})=e^{-rT}E^{Q}(e^{-\int^{T}_{0}\lambda_{s}^{Q}ds})
    \end{eqnarray*}
    
 \begin{proposition}
 Under the assumption that the market price of risk for the non-traded asset $Y$ under $Q$ is equal to zero, 
 
 \begin{eqnarray}
\lim_{\gamma \rightarrow 0} P_{0}^{b}(T)=\lim_{\gamma \rightarrow 0} P_{0}^{s}(T)=P^{ar}_{0}(T)\,\,\,\text{if}\,\, \mu=0
\end{eqnarray}

 \end{proposition}
 
 \begin{proof} When the risk aversion parameter $\gamma$ tends to zero, \cite{beche} proved the indifference price goes to the arbitrage free pricing under the minimal entropy martingale measure:
 
 \begin{eqnarray*}
\text{lim}_{\gamma \rightarrow 0} P_{0}^{b}(T)=\text{lim}_{\gamma \rightarrow 0} P_{0}^{s}(T)&=&E^{Q^{0}}(e^{-rT}1_{\left\{\tau>T\right\}})\\
&=&E^{P^{0}}(e^{-rT}1_{\left\{\tau>T\right\}})
\end{eqnarray*}
 If the excess return $\mu=0$, the minimal martingale measure $P^{0}$ coincides with the the investor's subjective measure $P$ and the result follows.
 
 \end{proof}   
%
%
%
%
\subsection{Solving the PDE}
  
With the change of time $t\rightarrow T-t$, the equation (\ref{M-function}) reduces  to:
\begin{eqnarray}\left\{\begin{array}{lcl}
-u_{t}+F_{1}(y)u_{yy}+F_{2}(y)u_{y}+F_{3}(y)u+F_{4}(y)u^{-\theta}=0\\ 
u(0,y)=1
\label{rod}
\end{array}\right.
\end{eqnarray}

with :
$$F_{1}(y)= \frac{1}{2}a(y)^{2}$$
$$F_{2}(y)=b(y)-\frac{\rho \mu}{\sigma}a(y)$$
$$F_{3}(y)= -\left(1-\rho^{2}\right)\left(\frac{\mu^{2}}{2 \sigma^{2}}+\lambda(y)\right)$$
$$F_{4}(y)=\left(1-\rho^{2}\right)\lambda(y)$$

\subsubsection{Finite Differencing of PDE}
Our aim is to solve (\ref{M-function})-(\ref{M2-function}) by implicit finite difference schemes. To this end, we divide the interval $[y_{min},y_{max}]$ into the sub intervals
$$y_{min}=y_{0}<y_{1}...<y_{N+1}=y_{max}$$
and we assume for convenience for the mesh-points $\left\{y_{j}\right\}_{{j=0}}^{{N+1}}$ are equidistant that is,\\

$y_{j}=y_{j-1}+\Delta y$, $j=1,...,N+1$     ($\Delta y=\frac{y_{max}-y_{min}}{N+1}$).\\

Furthermore, we divide the interval $[0,T]$ into $M+1$ equal sub-intervals
$$0=t_{0}<t_{1}...<t_{M+1}=T$$
where\\

$t_{i}=t_{i-1}+\Delta t$, $i=1,...,M+1$          ($\Delta t=\frac{T}{M+1}$)\\

Let $u(t_{i+1}, y_{j})=u_{j}^{i+1}$, the essence of the finite difference approach lies in replacing the derivatives in (\ref{M-function}) by divided differences at the mesh-points $(t_{i}, y_{j})$. The following equations are used to approximate the first and second order derivatives:\\
$$\frac{\partial u}{\partial t}=\frac{u_{j}^{i+1}-u_{j}^{i}}{\Delta t}$$
$$\frac{\partial u}{\partial y}=\frac{u_{j+1}^{i+1}-u_{j-1}^{i+1}}{2\Delta y}$$
$$\frac{\partial^{2} u}{\partial y^{2}}=\frac{u_{j+1}^{i+1}-2u_{j}^{i+1}+u_{j-1}^{i+1}}{(\Delta y)^{2}}$$
To avoid the difficulty of the nonlinear term, the following first order approximation is used
$$(u_{j}^{i+1})^{-\theta}=(1+\theta)(u_{j}^{i})^{-\theta}- \theta (u_{j}^{i+1})(u_{j}^{i})^{- \theta-1}.$$

Applying the above approximations to (\ref{rod}), we get
$$\alpha_{j} u_{j-1}^{i+1}+\beta_{j} u_{j}^{i+1}+\nu_{j} u_{j+1}^{i+1}=-u_{j}^{i}-\psi_{j} (u_{j}^{i})^{-\theta},$$ 
with
\begin{eqnarray*}
\alpha_{j}&=&\frac{\Delta t F_{1,j}}{\Delta y^{2}}-\frac{\Delta t F_{2,j}}{2\Delta y},\\
\beta_{j}&=&-1-\frac{2 \Delta t F_{1,j}}{\Delta y^{2}}+\Delta t F_{3,j}-\theta \Delta t F_{4,j}(u_{j}^{i})^{- \theta-1},\\
\nu_{j}&=&{\frac{\Delta t  F_{1,j}}{\Delta y^{2}}}+\frac{\Delta t F_{2,j}}{2\Delta y},\\
\psi_{j}&=&(1+\theta)\Delta t F_{4,j}.
\end{eqnarray*}
Thus we obtain the linear system $Ru^{i+1}=d^{i}$, $0\leq i\leq M$ where $R$ is formed matrix using the coefficients $\alpha_{j}$, $\beta_{j}$, $\nu_{j}$ and $d^{i}$ is a vector with $d_{j}^{i}=-u_{j}^{i}-\psi_{j} (u_{j}^{i})^{-\theta}$. The calculations are similar for $w^{b}$ and $w^{c}$, the only changes are $F_{4}(y)=e^{\gamma c}\left(1-\rho^{2}\right)\lambda(y)$ for $w^{b}$ and 
$F_{4}(y)=e^{-\gamma c}\left(1-\rho^{2}\right)\lambda(y)$ for $w^{c}$. \\

\subsubsection{Choice of parameters and Boundaries}    

The choice of the appropriate level of risk aversion $\gamma$ ($>0$) is important in the indifference pricing within an exponential utility function. If a high value is chosen, the investor will be too risk averse and therefore the results of the pricing will converge to the super-replication's prices. On the other hand if the value of risk aversion is too small, the decisions of the investor will not differ appreciably from the decisions of a risk neutral investor. 
In \cite{turkey}, the authors examine the risk attitudes of farmers in the Lower Seyhan Plain of Turkey. While some variation by utility function exists in the classification of the sampled farmers into risk averse and risk preferring categories, the overwhelming evidence is that the sampled farmers are risk averse. The study also found for the exponential utility function, the estimated values of $\gamma$ $\in$ $\left[0.0375;\, 0.4920\right]$ with a mean value of $0.11$.
 We use in our study different values of risk aversion coefficients such as $\gamma=0.01$, $\gamma=0.2$ and $\gamma=0.7$. We choose as \cite{SZ} $\mu=0.09$, $\sigma=0.15$, $r=0.03$. We work with negative correlation coefficients $\rho\in \left\{-0.90,-0.5,-0.20,-0.05\right\}$ in order to specify that the intensity tends to rise when the stock price falls. We define $\lambda_{t}=Y_{t}$ with $Y_{t}$ follows CIR dynamics defined below. 

\paragraph{Cox-Ingersoll-Ross intensities}

In this model, the intensity follows the stochastic differential equation 

\begin{eqnarray*}
d\lambda_{t}= \alpha (\bar{\lambda}-\lambda_{t})dt+\phi\sqrt{\lambda_{t}}dW_{t}^{3}
\end{eqnarray*}

where $\alpha$, $\bar{\lambda}$ and $\phi$ are positive constants and $dW_{t}^{3}=\rho dW_{t}^{1}+\rho'dW_{t}^{2}$. When we impose the condition $2\alpha\bar{\lambda}>\phi^{2}$ then the intensity $\lambda$ is always positive, otherwise we can only guarantee that it is non-negative (with a positive probability to terminate to zero). In fact, if the intensity approaches zero then the volatility $\phi\sqrt{\lambda_{t}}$ approaches zero cancelling the effect of the randomness, so the intensity rate remains always non-negative. We fix $\lambda_{min}=0$ and $\lambda_{max}=1$.\\

When $\lambda_{t}=1$, the agent will follow the sub-optimal policy in investing exclusively in the default-free bank account ($\pi_{t}=0$). \\ 

$M(t,x,\lambda=1)=-e^{-\gamma x}$ and therefore $u(t,\lambda_{max})=1$. 


\begin{eqnarray*}
H^{b}(t,x,\lambda=1)&=&E\left(-e^{-\gamma (x+c)}1_{\{\tau_{t} >T\}}+(-e^{-\gamma x})1_{\{\tau_{t} \leq T\}} \vert X_{t}=x, \lambda_{t}=1 \right)\\
&=& -e^{-\gamma x}+(e^{-\gamma x}-e^{-\gamma (x+c)}) P(\tau_{t} >T \vert \lambda_{t}=1)\\
P(\tau_{t} >T \vert \lambda_{t}=1)&=&1_{\{\tau>t\}}E(e^{-\int_{t}^{T} \lambda_{s} ds} \vert \lambda_{t}=1)\\
&=&1_{\{\tau>t\}}A(t,T)e^{-B(t,T)}
\label{bid5}
\end{eqnarray*}

 with
 \begin{eqnarray*}
 B(t,T)&=&\frac{2 (e^{\xi (T-t)}-1)}{2 \xi+(\alpha+\xi)(e^{\xi (T-t)}-1)}\\
 A(t,T)&=&\left\{\frac{2\xi e^{(\alpha+\xi)(\frac{(T-t)}{2})}}{2 \xi+(\alpha+\xi)(e^{\xi (T-t)}-1)}\right\}^{ \frac{2\alpha   \bar{\lambda}}{\phi^{2}}}\\
 \xi&=&\sqrt{\alpha^{2}+2\phi^{2}}
 \end{eqnarray*}
 
 
 
Then for the buyer,\\

 $w^{b}(t,\lambda_{max})=\left(e^{\gamma c}-(e^{\gamma c}-1)1_{\{\tau>t\}}A(t,T)e^{-B(t,T)} \right)^{1-\rho^{2}}$\\

Similarly for the seller,\\

$w^{s}(t,\lambda_{max})=\left(e^{-\gamma c}-(e^{-\gamma c}-1)1_{\{\tau>t\}}A(t,T)e^{-B(t,T)} \right)^{1-\rho^{2}}$\\

See \cite{CIR} for the derivation of $A(t,T)$ and $B(t,T)$.\\

When $\lambda_{t}=0$, (\ref{M-function}) gives the following equation

\begin{eqnarray}\left\{\begin{array}{lcl}
\frac{\partial u}{\partial t}+\alpha \bar{\lambda} \frac{\partial u}{\partial \lambda}-(1-\rho^{2})\frac{\mu^{2}}{2 \sigma^{2}}u=0\\
u(T,0)=1
\end{array}\right.
\label{hj}
\end{eqnarray}


 The equation at hand is a linear Hamilton-Jacobi equation with the hamiltonian is a function of $\frac{\partial u}{\partial \lambda}$ and $u$. The solution of (\ref{hj}) is 
 
 \begin{eqnarray*}
u(t,0)=e^{-(1-\rho^{2})\frac{\mu^{2}}{2 \sigma^{2}}(T-t)}
\end{eqnarray*}
 
 Similarly, for the buyer and seller
 
 \begin{eqnarray*}
w^{b}(t,0)=w^{s}(t,0)=e^{-(1-\rho^{2})\frac{\mu^{2}}{2 \sigma^{2}}(T-t)}
\end{eqnarray*}

For the implementation of the model, we use the following estimated values for CIR-intensity given in \cite{longstaff}: $\alpha=0.20$, $\bar{\lambda}=0.06$, $\phi=0.03$.

\paragraph{Numerical Results and Comments}
           
 In this paragraph, we present the results of the indifference pricing of the defaultable bond solved by the finite difference method. The yield-spread curve which reflects the relation between yield spreads and the time to maturity is  plotted and analyzed. The yield-spread is the difference between the yield on the defaultable bond and the yield on the risk free bond (in our situation $r$). The yield spread is the indication of the risk premium required for investing in the risky bond. When it is high, this means that the investors require a higher risk premium, so the bond is more risky. Conversely when it is small, this means that the bond is riskless. 

In the figures \ref{fig1} and \ref{fig2}, we present the spread curves for the buyer and the seller respectively for various risk aversion coefficients given the intensity $\lambda=0$, $\lambda=0.02$, $\lambda=0.12$ and $\lambda=0.5$. When $\lambda\neq0$ the short term limit of the yield spread for the buyer and seller is nonzero, reflecting the presence of non-predictable defaults. Even if the intensity and the risk aversion parameter are small, the investor requires at short term a risk premium. We can remark for the maturity $T=0$, $\lambda=0.02$ and $\gamma=0.01$, the bid and ask spreads are equal to $0.045$. When the default intensity increases, the bid and ask spread curves are in fact high in order to compensate the additional risk; but the shape of the curves changes depending if it is a buyer or a seller. An experiment on many values of intensity shows the following shapes for the buyer's spread curve in the figure \ref{fig1},

\begin{itemize}
	\item $0<\lambda\leq 0.06$, the spread curves are upward sloping reflecting a low default risk at short term, whereas the forecast of the credit quality of the firm over longer maturities is less certain. This shape is illustrated for $\lambda=0.02$.
	\item $0.06 < \lambda \leq 0.15$, the spread curves are $S$ shaped that is the buyer expects that the credit quality of the firm will increase until a certain maturity and after that it will be low for longer maturity. This particular shape is shown for $\lambda=0.12$.
	\item $0.15 < \lambda \leq 1$, the spread curves are downward sloping reflecting a high default probability of the firm at short term; once the firm has survived a certain period of time without a default, it faces a lower default probability in the long term. The shape is shown for $\lambda=0.5$.
\end{itemize}
For the seller's spread curves, we have two kinds of shape in the figure \ref{fig2}, 
 
 \begin{itemize}
	\item $0<\lambda\leq 0.25$, the curve is upward sloping and is shown for $\lambda=0.02$.
	\item $0.25<\lambda\leq 1$, the curves are humped meaning that the seller expects to offer a high risk premium until a certain maturity. Once the firm doesn't default until this reference maturity, the ask yield-spread decreases. We show the humped curves for $\lambda=0.50$ and $\lambda=0.90$.
	\end{itemize}
	For $\lambda=0$, both buyer and seller's spreads are closed to zero whatever the maturity since there is no default probability. The impact of the risk aversion $\gamma$ on the yield-spread is also analyzed. For the buyer, we found a positive relationship between the spread and the risk aversion coefficient in contrast to the negative relationship for the seller. It means that if the buyer is more risk averse, he will require a high risk premium for investing in the defaultable bond. In contrast, a more risk averse seller will offer a low risk premium for selling the defaultable bond. This result is also a consequence of the monotonicity of the indifference price as a function of risk aversion parameter. For more details see \cite{sij}. \\
	
	We analyze in figures \ref{fig3} and \ref{fig4}, the impact of the correlation coefficient $\rho$ on the spread curves. At the maturity $T=0$, the correlation between the intensity and the stock price has no impact on the yield-spread for both buyer and seller. But for longer maturities, $\rho$ (in absolute value) increases the yield-spreads for the investors. At time $T=50$, we can clearly remark the impact of $\rho$ on the bid and ask's spread curves. When $\left|\rho\right|=1$, the intensity and the stock price are perfectly linearly dependent and the unique source of uncertainty is the stock price which is a traded asset. This leads to the complete market. In the figures \ref{fig5} and \ref{fig6}, we show the convergence of the indifference spread to the classical spread for $\mu=0$ and $\gamma=0.0001$

\begin{figure}[H]
\begin{minipage}[b]{.46\linewidth}
\begin{center}
\includegraphics[width=2.7in]{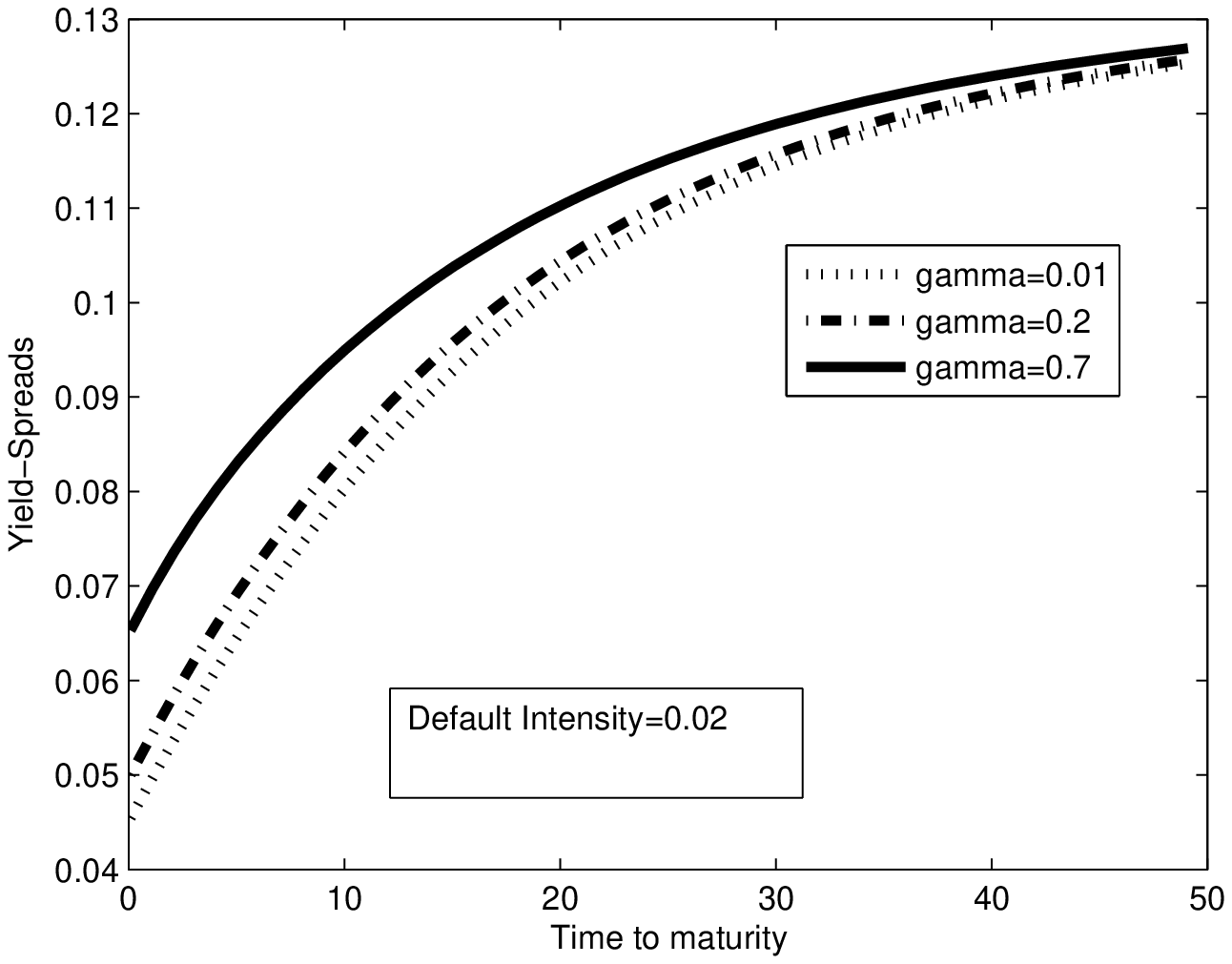}
\includegraphics[width=2.7in]{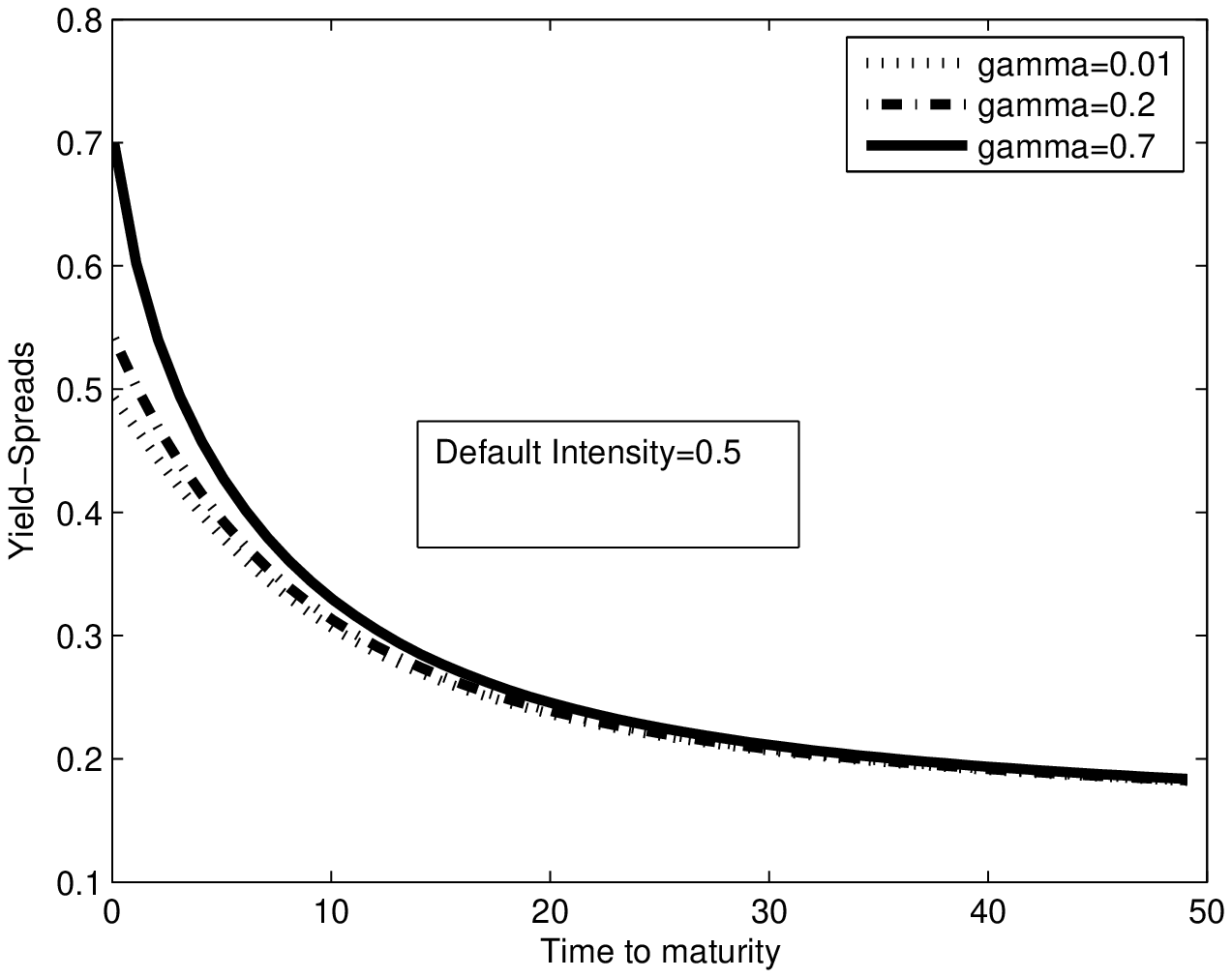}
\end{center}
\end{minipage}
\begin{minipage}[b]{.46\linewidth}
\begin{center}
\includegraphics[width=2.7in]{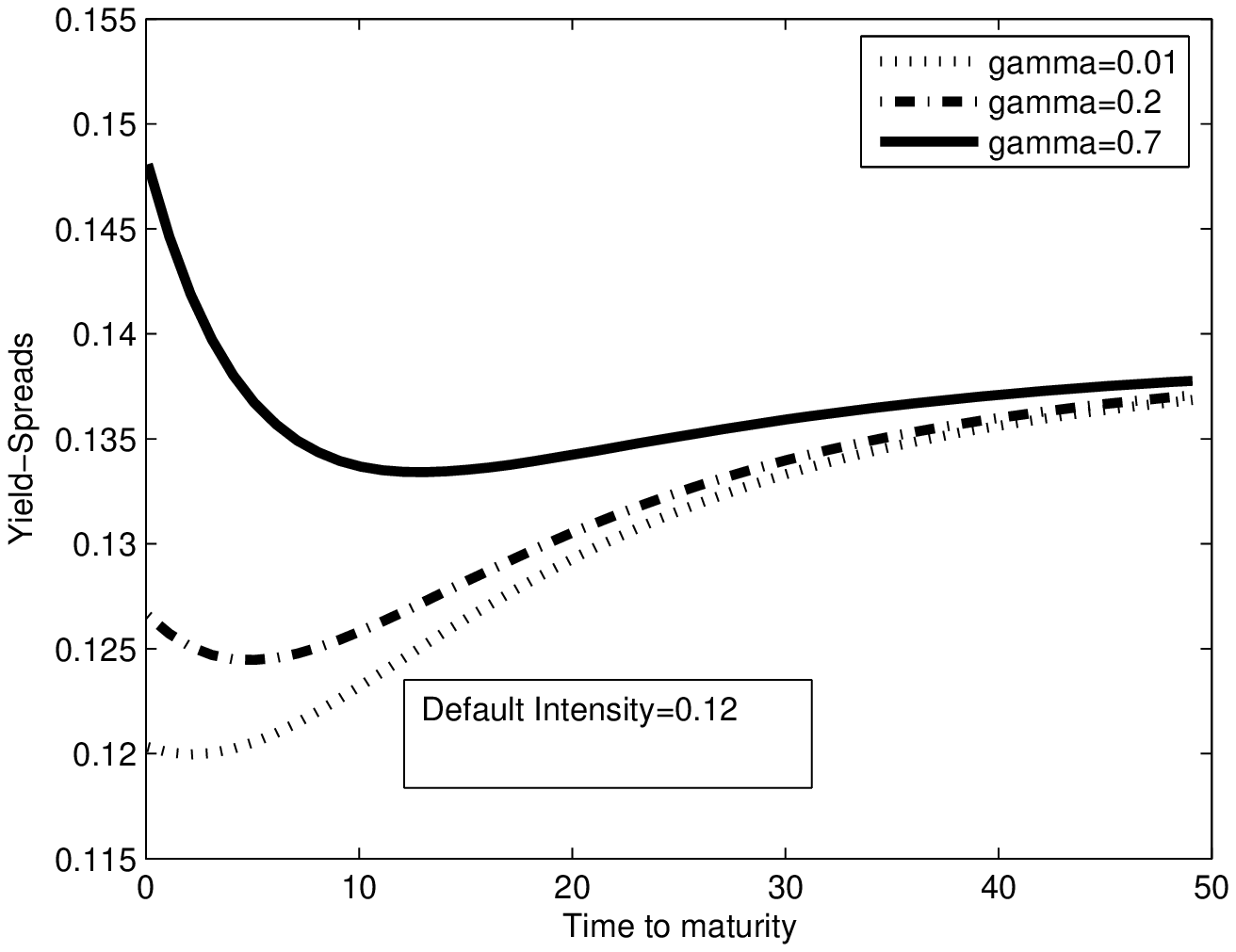}
\includegraphics[width=2.7in]{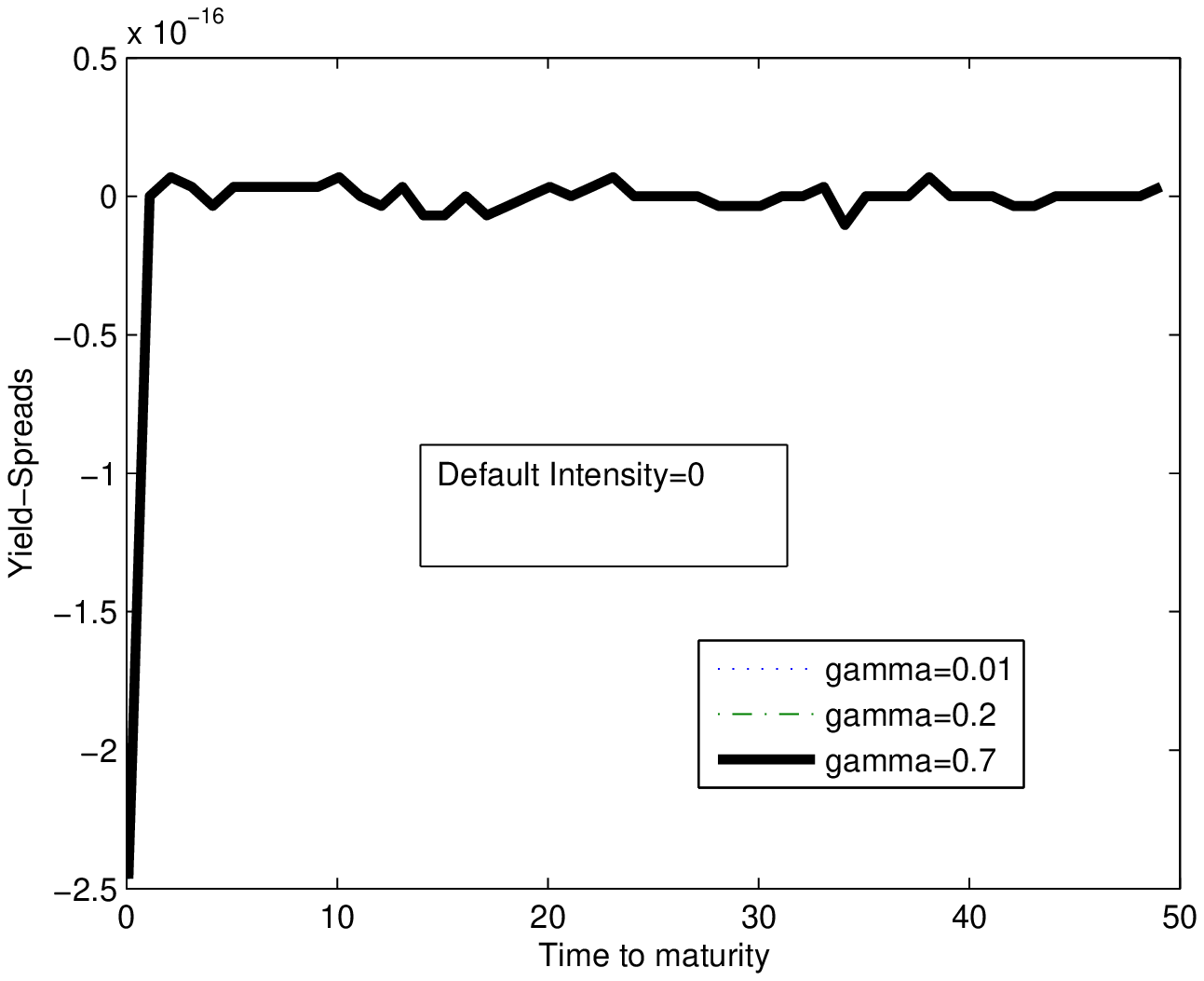}
\end{center}
\end{minipage}
\caption{Buyer's yield-Spreads for CIR intensities, $\rho=-0.10$ }
\label{fig1}
\end{figure}

\begin{figure}[H]
\begin{minipage}[b]{.46\linewidth}
\begin{center}
\includegraphics[width=2.7in]{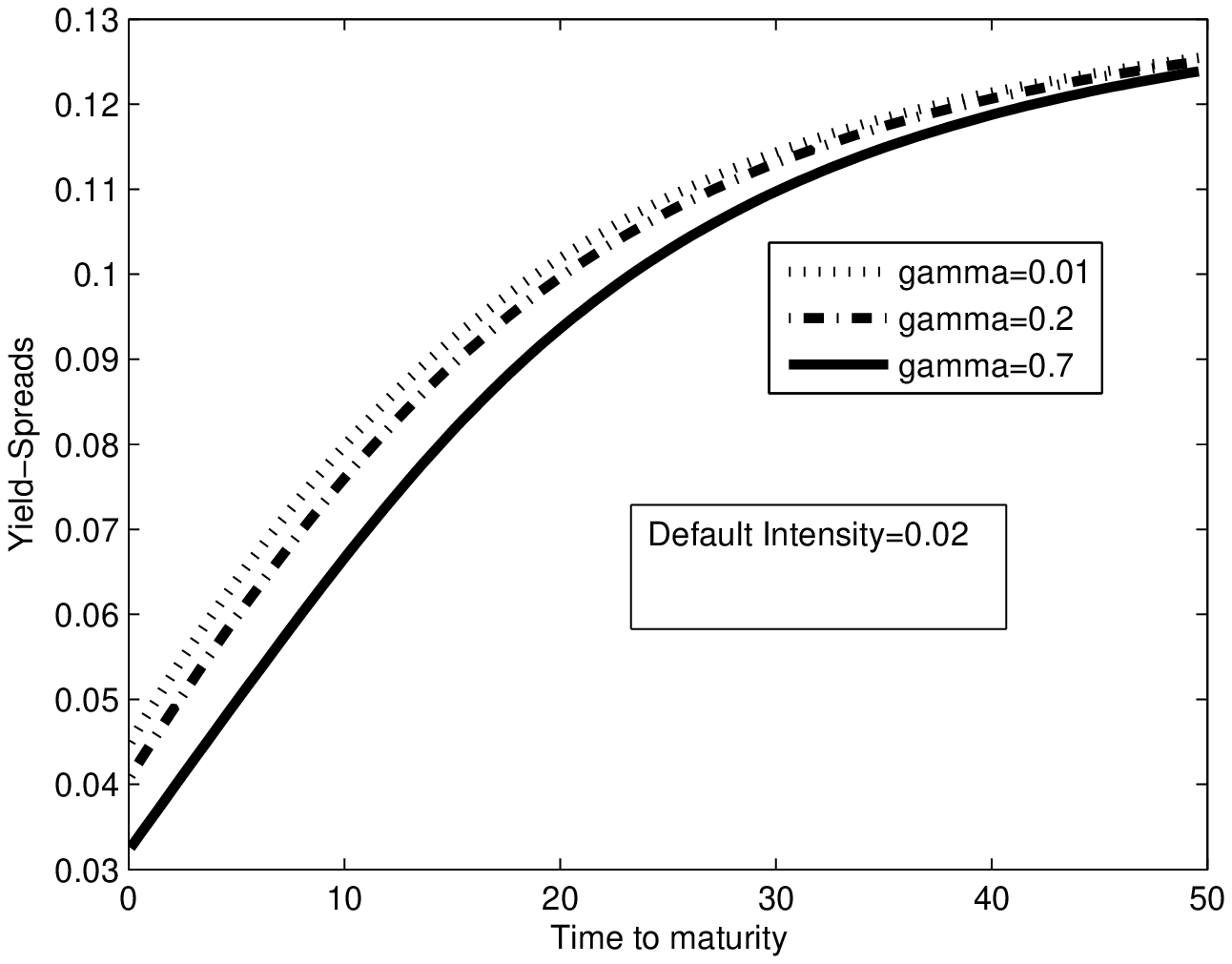}
\includegraphics[width=2.7in]{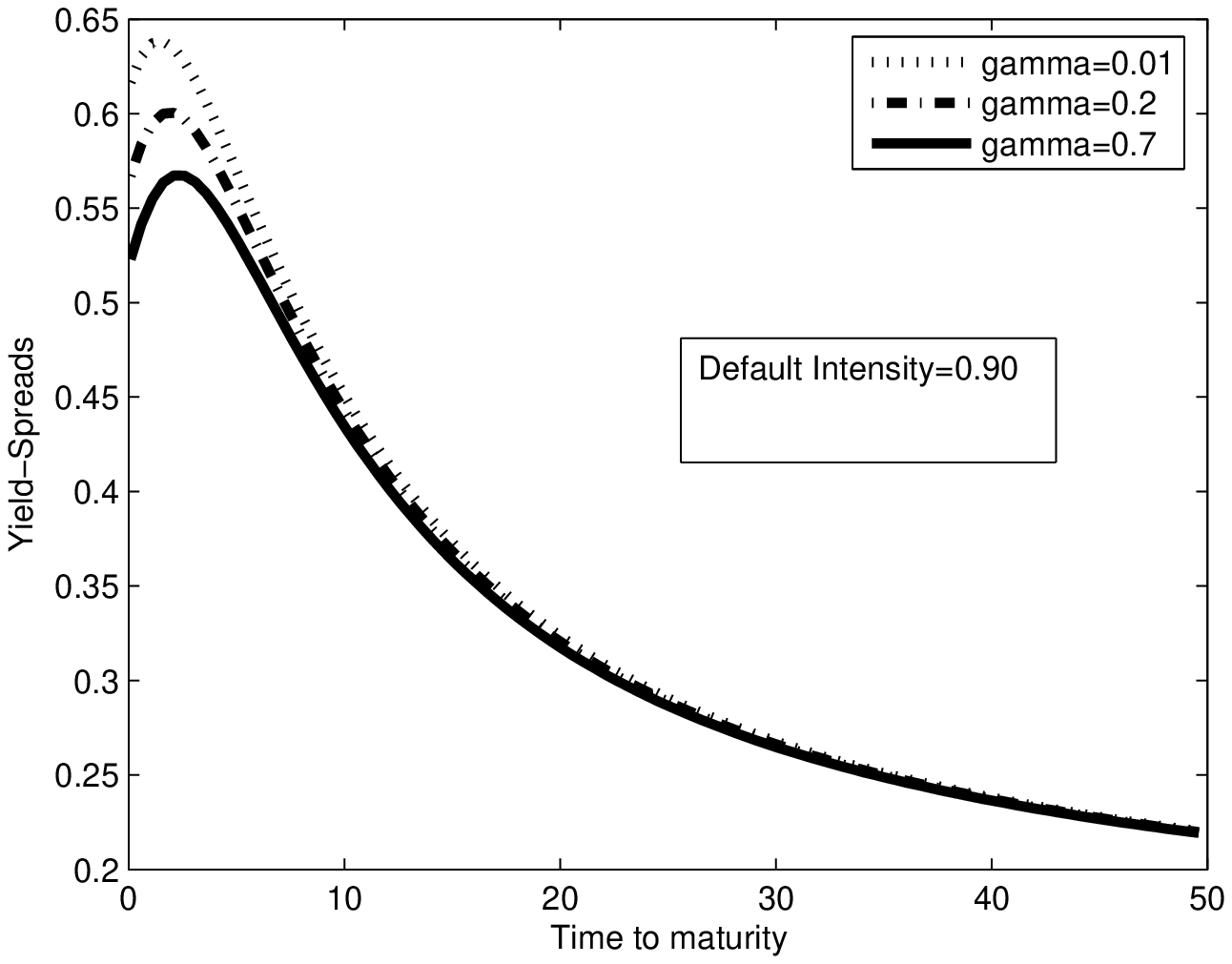}
\end{center}
\end{minipage}
\begin{minipage}[b]{.46\linewidth}
\begin{center}
\includegraphics[width=2.7in]{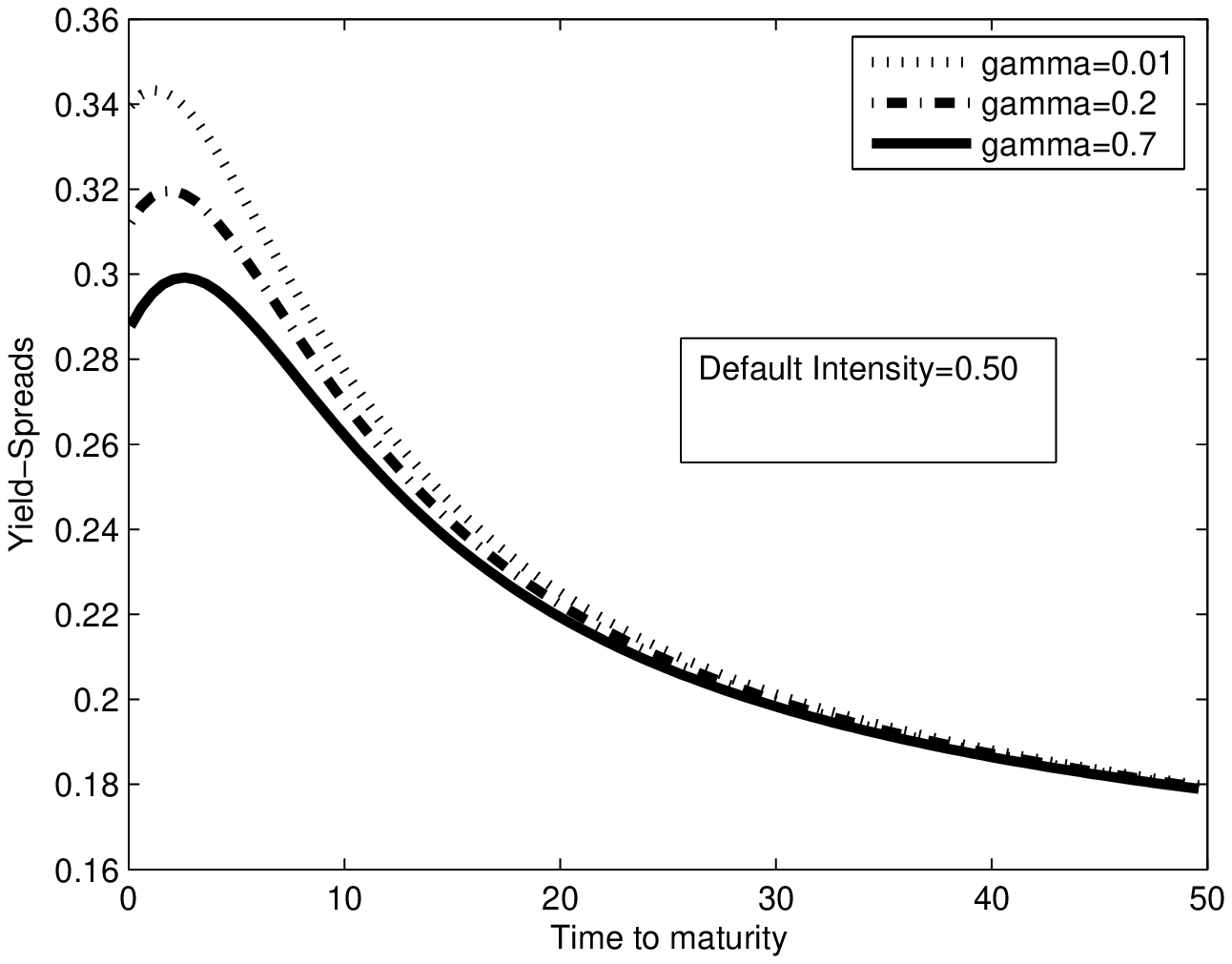}
\includegraphics[width=2.7in]{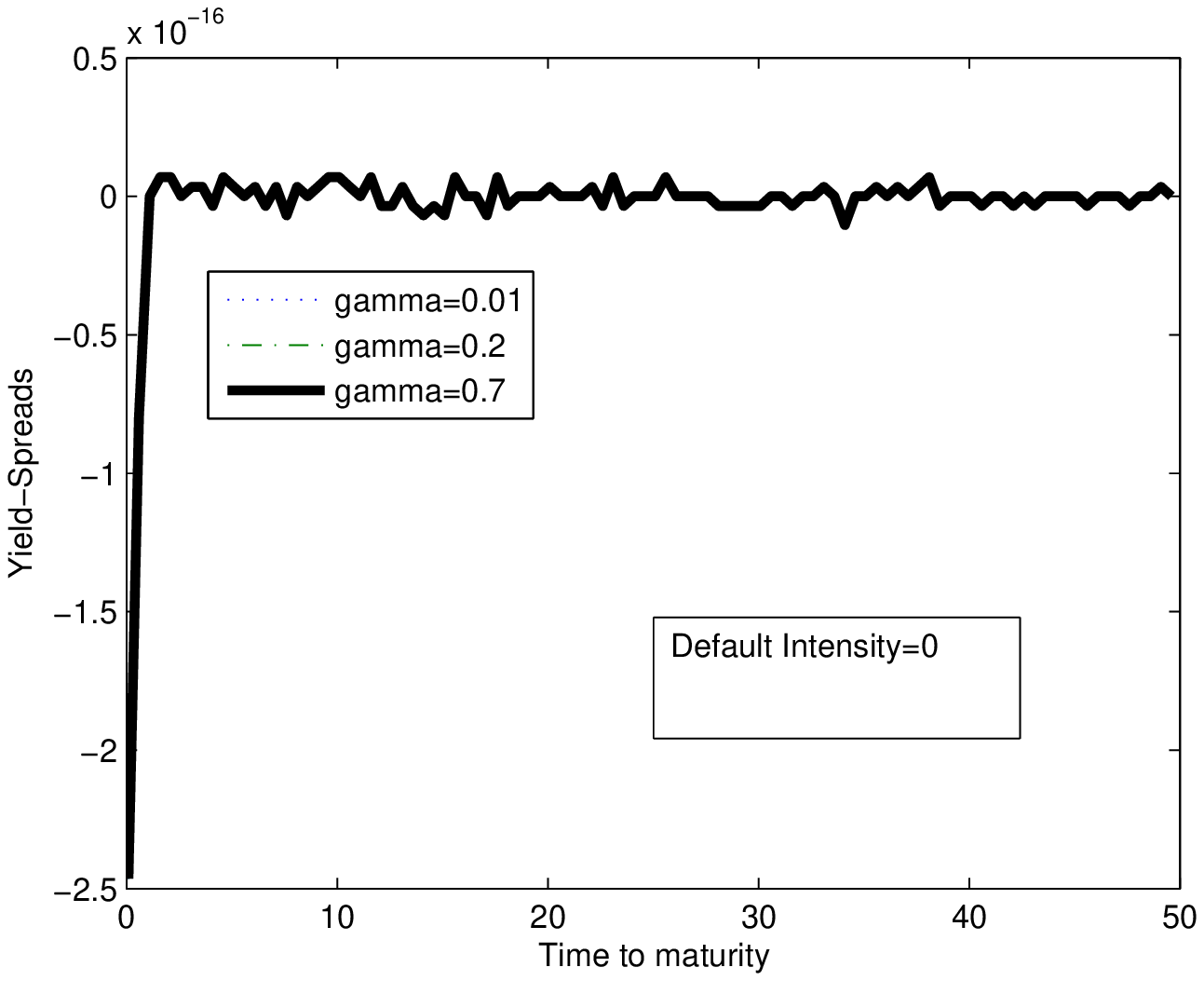}
\end{center}
\end{minipage}
\caption{Seller's yield-Spreads for CIR intensities, $\rho=-0.10$}
\label{fig2}
\end{figure}

\begin{figure}[H]
\begin{minipage}[b]{.46\linewidth}
\begin{center}
\includegraphics[width=2.7in]{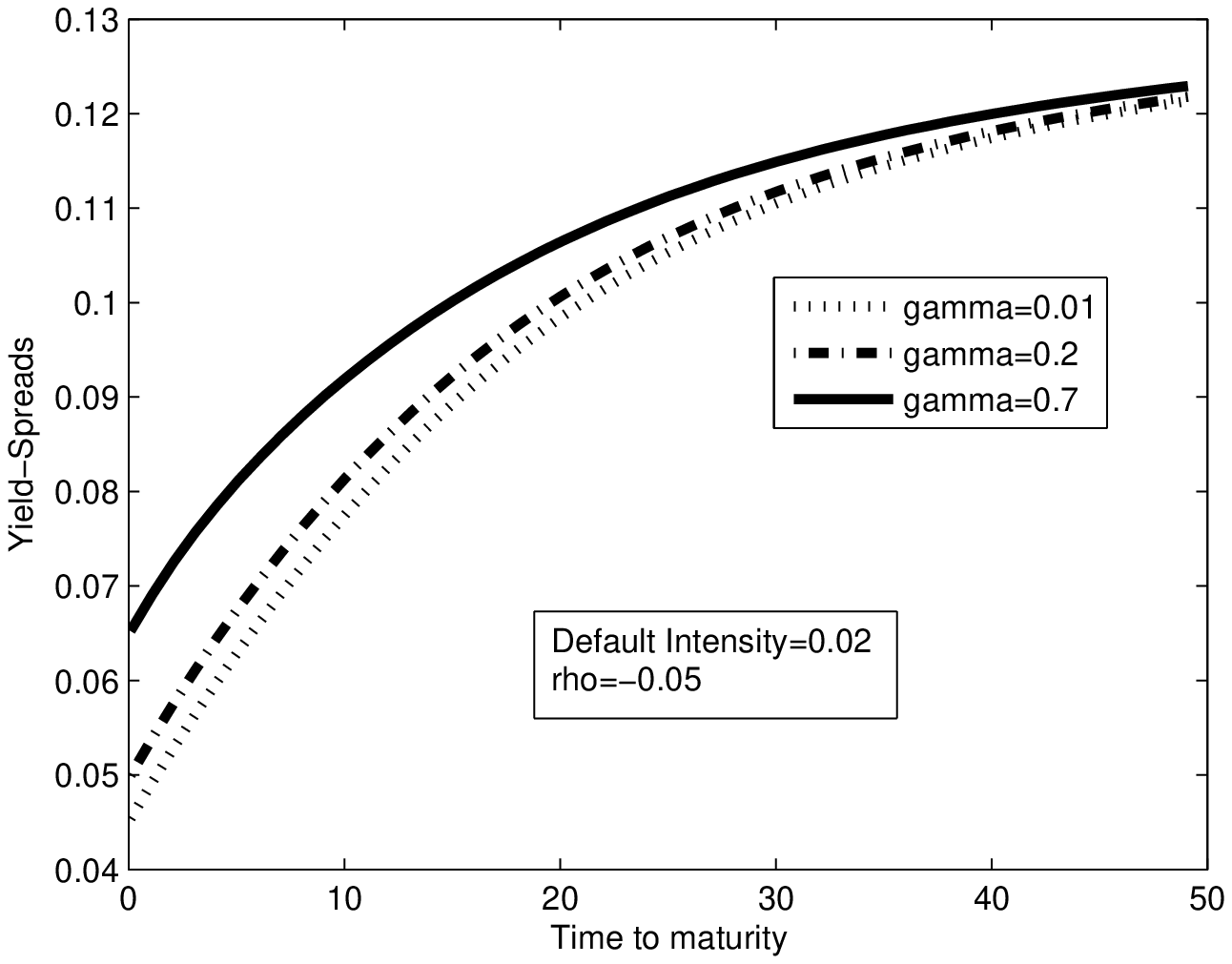}
\includegraphics[width=2.7in]{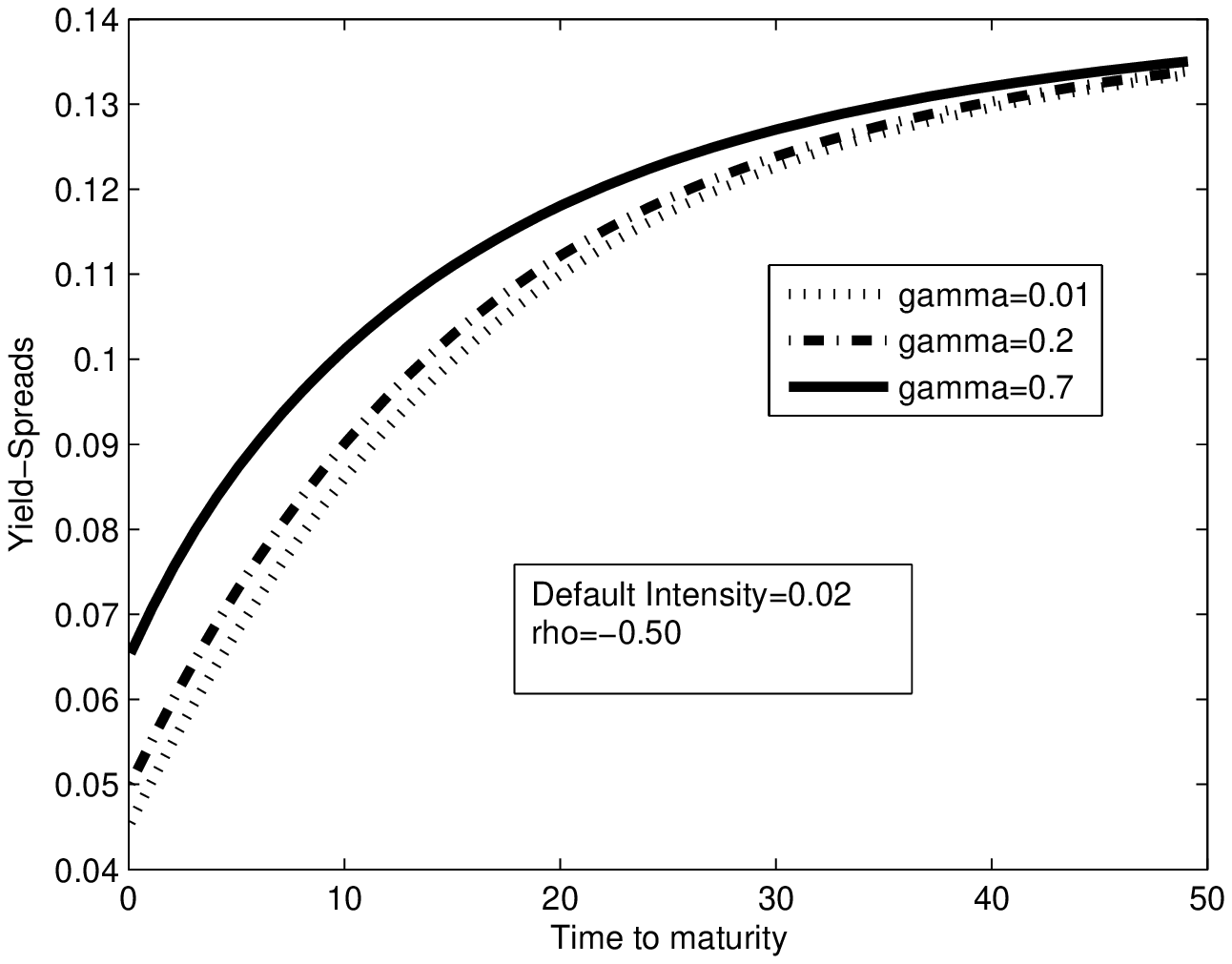}
\end{center}
\end{minipage}
\begin{minipage}[b]{.46\linewidth}
\begin{center}
\includegraphics[width=2.7in]{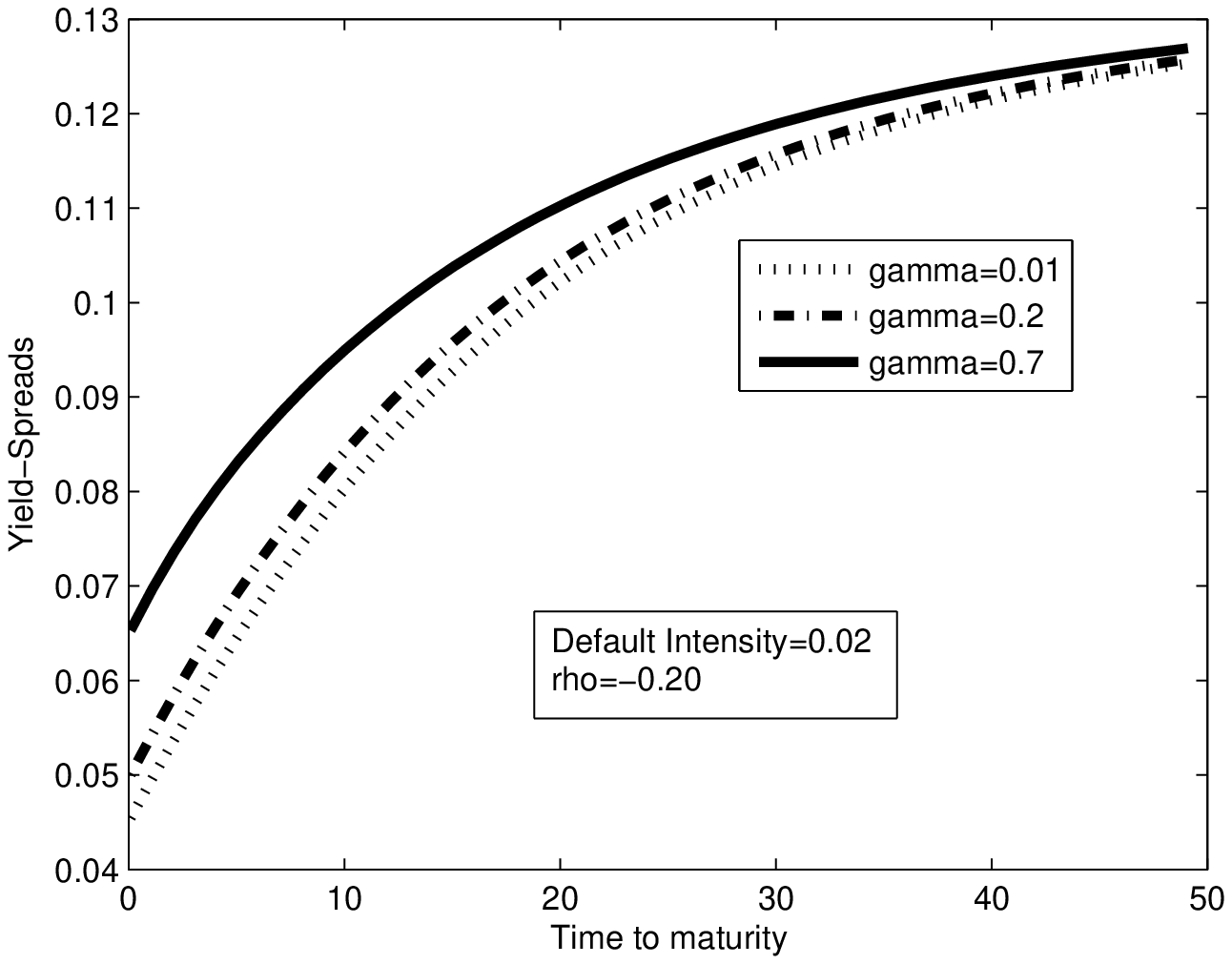}
\includegraphics[width=2.7in]{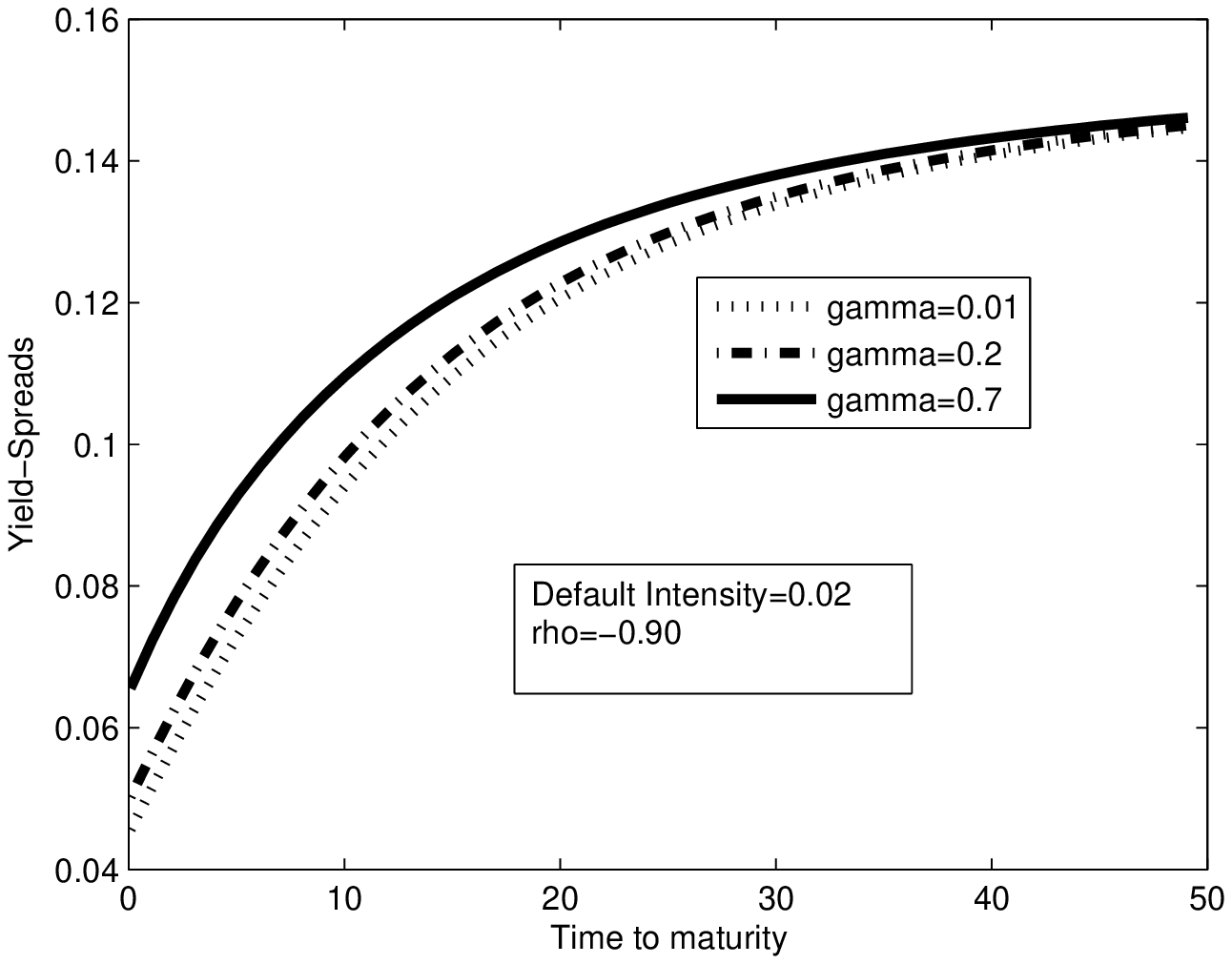}
\end{center}
\end{minipage}
\caption{Impact of $\rho$ on the bid spread-curves}
\label{fig3}
\end{figure}

\begin{figure}[H]
\begin{minipage}[b]{.46\linewidth}
\begin{center}
\includegraphics[width=2.7in]{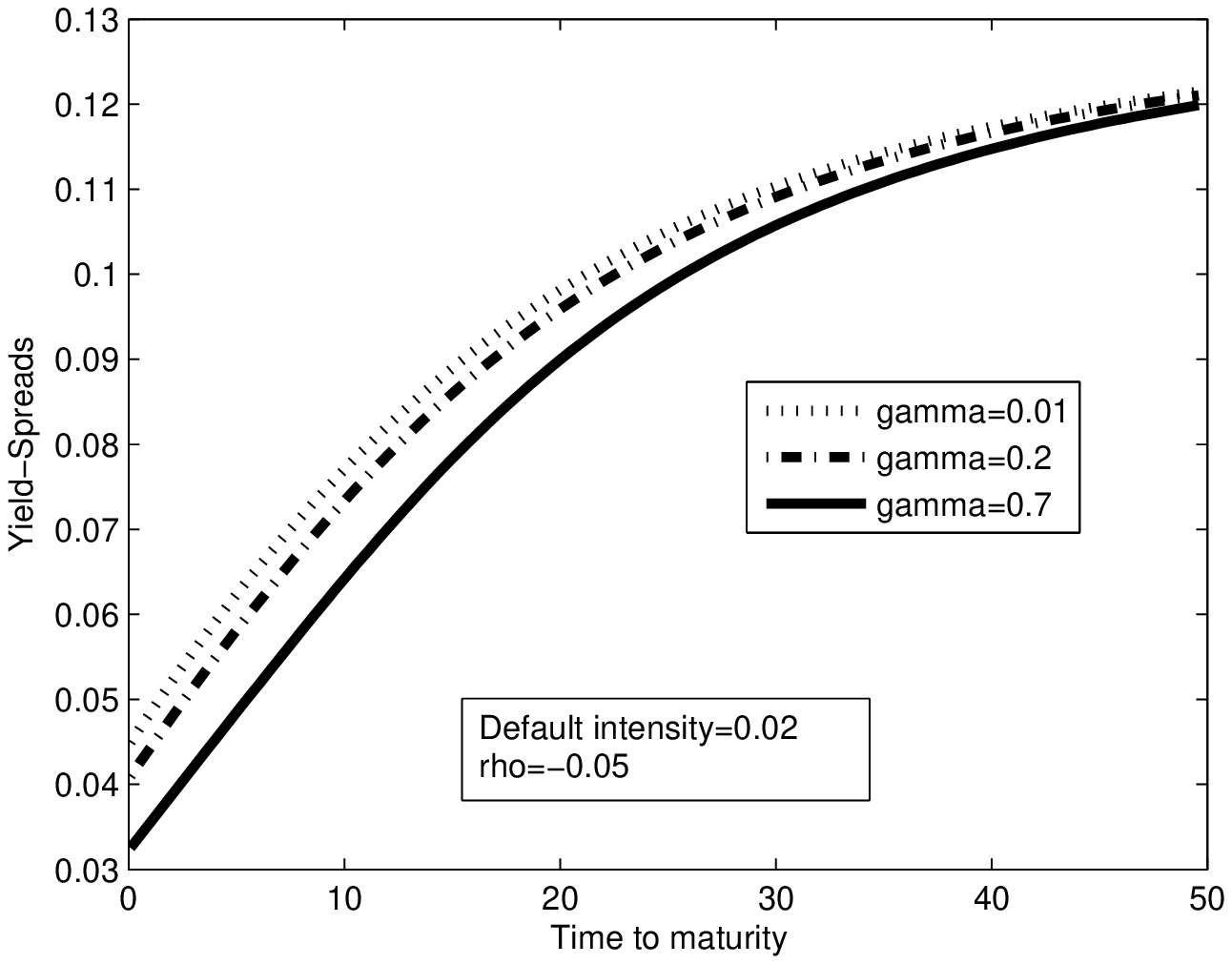}
\includegraphics[width=2.7in]{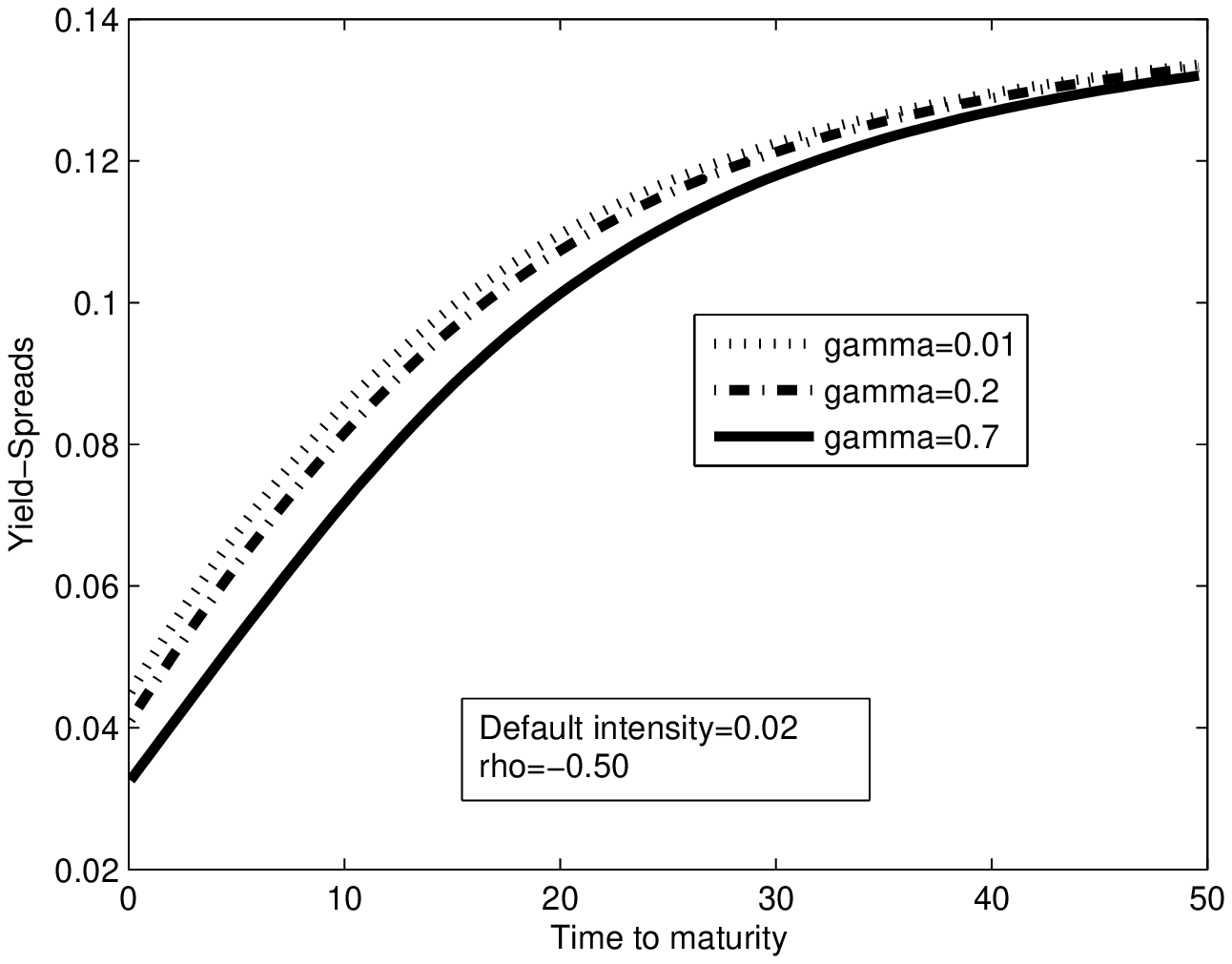}
\end{center}
\end{minipage}
\begin{minipage}[b]{.46\linewidth}
\begin{center}
\includegraphics[width=2.7in]{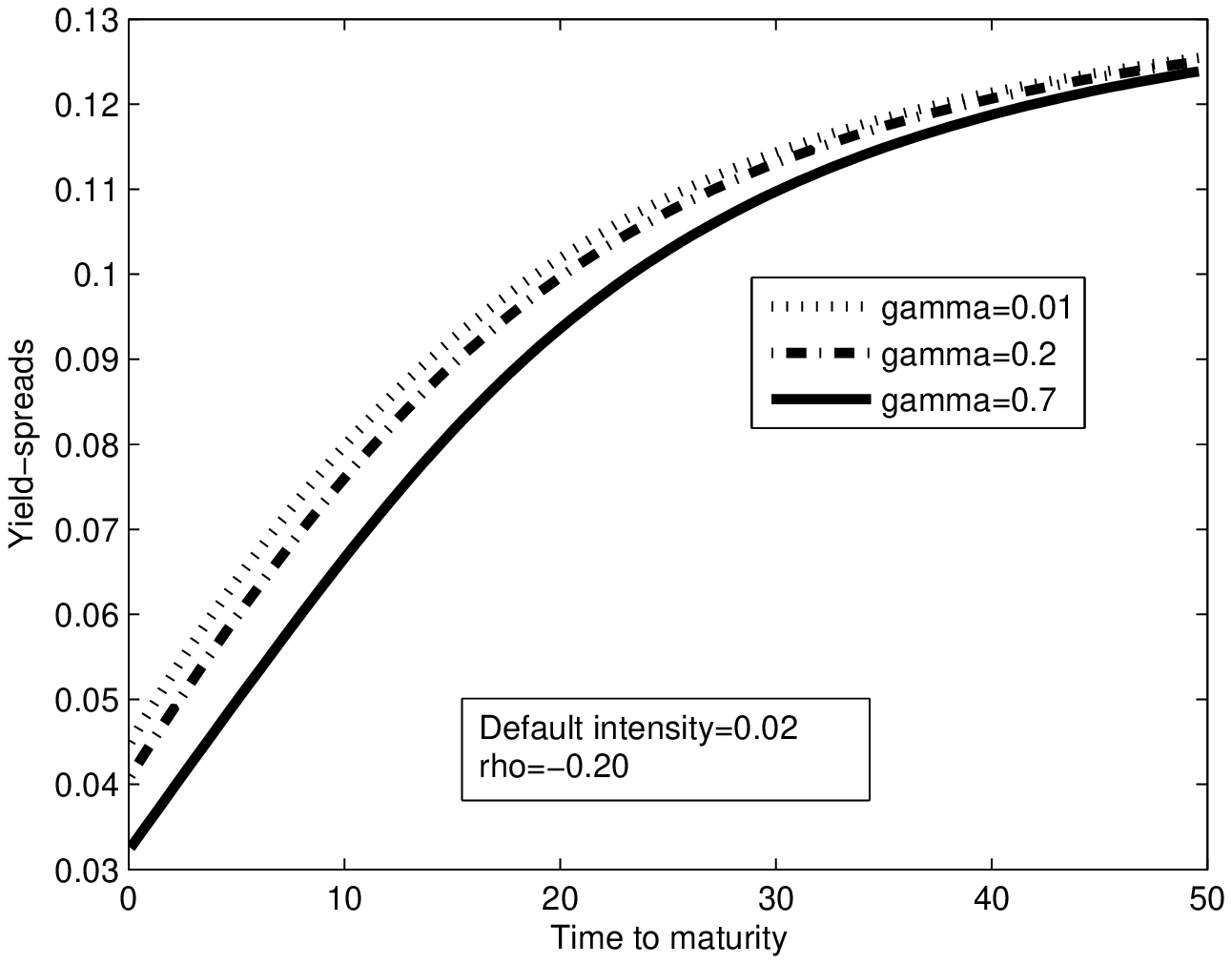}
\includegraphics[width=2.7in]{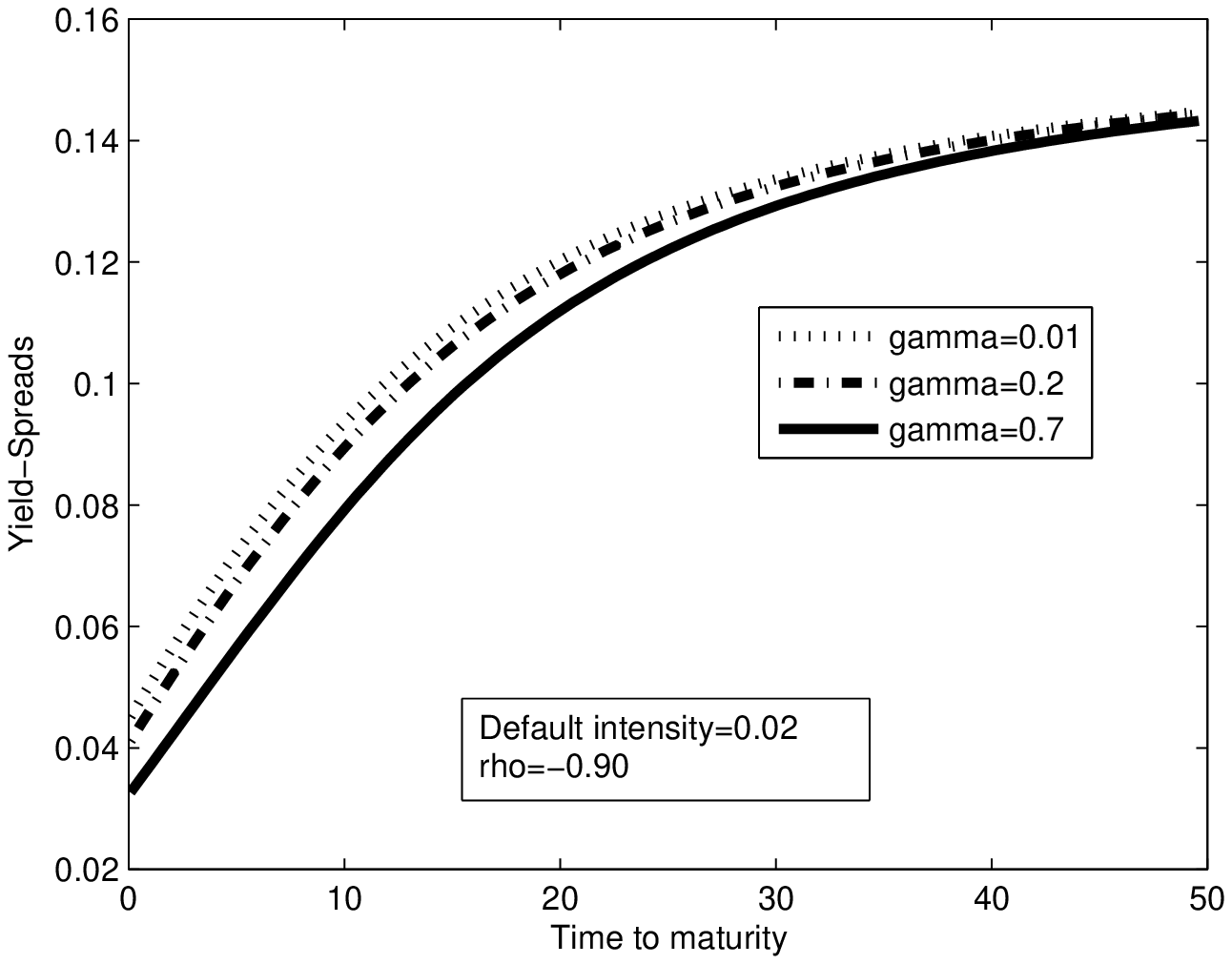}
\end{center}
\end{minipage}
\caption{Impact of $\rho$ on the ask spread-curves}
\label{fig4}
\end{figure}

\begin{figure}[H]
\begin{center}
\rotatebox{0}{\includegraphics[width=2.7in]{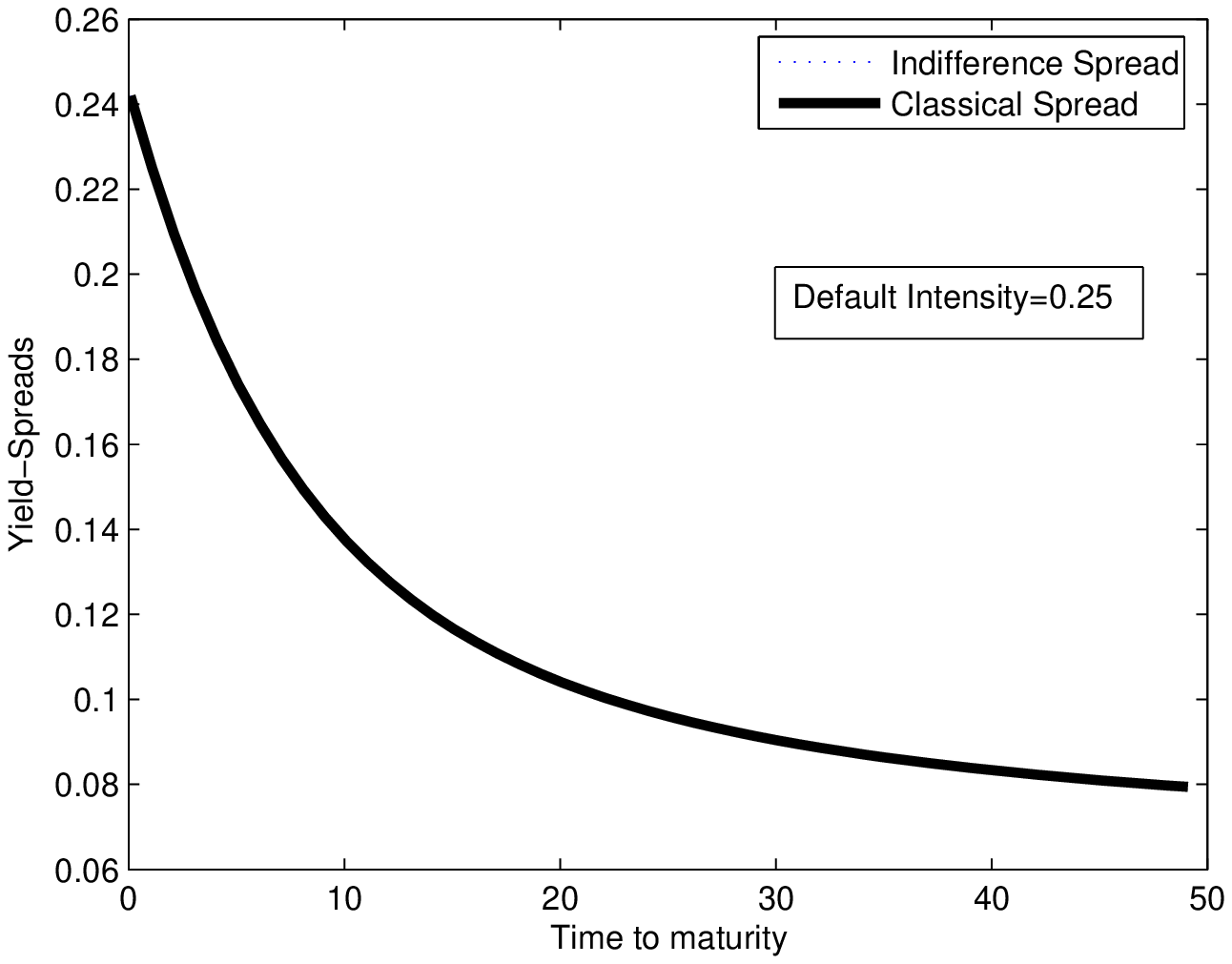}}
\caption{Indifference Spreads and Classical Spreads for buyer, $\mu=0$, $\gamma=0.0001$}
\label{fig5}
\end{center}
\end{figure}

\begin{figure}[H]
\begin{center}
\rotatebox{0}{\includegraphics[width=2.7in]{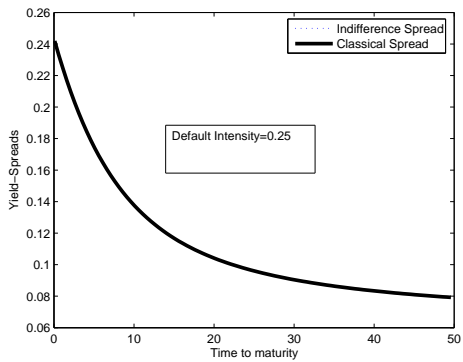}}
\caption{Indifference Spreads and Classical Spreads for seller, $\mu=0$, $\gamma=0.0001$}
\label{fig6}
\end{center}
\end{figure}

\section{Conclusion}
\label{conclusion}

The classical reduced models for credit risk were modified by embedding them in the framework of the utility based valuation. The reason of doing this, as mentioned in the introduction, is that the default intensity is a source of uncertainty that is not traded and the equivalent martingale measure is not unique. This leads to work with the utility indifference valuation for defaultable bonds. In our model, besides the uncertainty of the intensity, we introduce another correlated diffusion process, which comes from the firm's stock price to drive the credit risk. The behavior of the spread curve in our model is different from the classical reduced models due to the introduction of the diffusion process. Two news parameters are introduced in the model, namely the risk aversion parameter of the investor and the correlation coefficient between the intensity and the stock price, which results in the nonlinearity of the pricing rule. We derived the Hamilton-Jacobi-Bellman equations (HJB) verified by the values functions, and reduced them to the nonlinear reaction-diffusion equations. The nonlinear partial differential equations were solved by the implicit finite difference scheme. The effects of the intensity, the risk aversion parameter and the correlation coefficient on the spread curves when analyzed. Although the default intensity increases the bid and ask's yield spread, it changes the shape of the spread curves. For the buyer, the curves can be upward sloping, downward sloping and $S$ shaped depending of the values for the intensity. For the seller, we got only an upward sloping and humped spread curve. The other analysis reveal a positive relationship between the bid spread and the risk aversion coefficient and a negative relationship between the ask spread and the risk aversion coefficient. It reveals that the correlation coefficient increases the bid and ask 's spreads only for longer maturities. Another important topic, which is not considered in this paper is the asymptotic behavior of the spread curves within the stochastic intensity and the pricing of the Credit Default Swap. Both of these problems will be analyzed in future work.

\newpage
\bibliographystyle{apalike}
\bibliography{toto}

\end{document}